\documentclass[twocolumn,prl,amsmath,amssymb,aps,superscriptaddress,floatfix,preprintnumbers]{revtex4-1}
\usepackage{times}
\usepackage{graphicx}
\usepackage{amsmath, braket,amsfonts}
\usepackage{amssymb}
\usepackage{bm, color, ulem}
\begin{document}
\title{Robust fractional quantum Hall states and continuous quantum phase transitions in a half-filled bilayer graphene Landau level}
\author{A.A. Zibrov}
\affiliation{Department of Physics, University of California, Santa Barbara CA 93106 USA}
\author{C. Kometter}
\affiliation{Department of Physics, University of California, Santa Barbara CA 93106 USA}
\author{H. Zhou}\affiliation{Department of Physics, University of California, Santa Barbara CA 93106 USA}
\author{E.M. Spanton}\affiliation{California Nanosystems Institute, University of California at Santa Barbara, Santa Barbara, CA, 93106}
\author{T. Taniguchi}\affiliation{Advanced Materials Laboratory, National Institute for Materials Science, Tsukuba, Ibaraki 305-0044, Japan}
\author{K. Watanabe}
\affiliation{Advanced Materials Laboratory, National Institute for Materials Science, Tsukuba, Ibaraki 305-0044, Japan}
\author{M. P. Zaletel}
\affiliation{Department of Physics, Princeton University, Princeton, NJ 08544, USA}
\author{A.F. Young}
\affiliation{Department of Physics, University of California, Santa Barbara CA 93106 USA}
\date{\today}
\maketitle

\textbf{Nonabelian anyons offer the prospect of storing quantum information in a topological qubit protected from decoherence, with the degree of protection determined by the energy gap separating the topological vacuum from its low lying excitations\cite{kitaev_fault-tolerant_2003}.
Originally proposed to occur in semiconductor quantum wells in high magnetic fields\cite{nayak_non-abelian_2008}, experimental systems predicted to harbor nonabelian anyons now range from p-wave superfluids to superconducting systems with strong spin orbit coupling.
However, all of these systems are characterized by small energy gaps, and despite several decades of experimental work, definitive evidence for nonabelian anyons remains elusive.
Here, we report the observation of robust incompressible, even-denominator fractional quantum Hall (FQH) phases\cite{willett_observation_1987,ki_observation_2014,falson_even-denominator_2015} in a new generation of dual-gated, hexagonal boron nitride encapsulated bilayer graphene samples. Numerical simulations suggest that this state is in the Pfaffian phase\cite{moore_nonabelions_1991} and hosts nonabelian anyons.
The measured thermodynamic\cite{eisenstein_compressibility_1994} and transport energy gaps are several times larger than those observed in other systems\cite{cooper_observable_2009,nayak_non-abelian_2008}. Moreover, the unique electronic structure of bilayer graphene endows this platform with two new control parameters.
Tuning the magnetic field continuously changes the form of the effective electron interactions.
We find that the even-denominator gap changes non-monotonically, consistent with predictions that a transition between the Pfaffian phase and the composite Fermi liquid (CFL)\cite{papic_tunable_2011, metlitski_cooper_2015} occurs just beyond the experimentally explored magnetic field range.
The electric field, meanwhile, tunes crossings between Landau levels from different valleys \cite{liu_evolution_2011,falson_even-denominator_2015}.
Using a new capacitive technique to directly measure the valley polarization, we observe a continuous transition from an incompressible to a compressible phase at half-filling mediated by an unexpected incompressible, yet polarizable, intermediate phase. Conservation of the valley quantum number implies this phase is an electrical insulator with gapless neutral excitations.}

At high magnetic fields, two-dimensional electrons form flat bands, known as Landau levels (LLs). At finite charge density $n$, interactions drive the formation of ordered states depending on both the  LL filling ($\nu= 2\pi\ell_B^2 n$, where $\ell_B=\sqrt{\tfrac{\hbar}{eB}}$ is the magnetic length) as well as the spin and orbital structure of the LL wavefunctions. Of particular interest is the fate of the half-filled LL, which can be understood as a weakly interacting state of composite fermions (CFs)\cite{jain_composite-fermion_1989} consisting of one electron and two magnetic flux quanta.  Having bound part of the external magnetic field, the CFs experience an effective field $B_{\textrm{eff}}=B(1-2\nu)$. At $\nu = \tfrac{1}{2}$ this field vanishes and the CFs form an emergent Fermi surface\cite{halperin_theory_1993} that manifests in both microwave and transport experiments\cite{willett_experimental_1993,kang_how_1993}.  As in a conventional metal, the emergent Fermi surface can be unstable, depending on the strength and sign of the residual interactions between the CFs. Most intriguingly, CFs have been predicted to form the quantum Hall analog of a superconductor \cite{moore_nonabelions_1991,read_paired_2000} which, in a single component system, naturally has p-wave pairing symmetry and supports nonabelian, charge-$e/4$ quasiparticle excitations in an incompressible liquid.
Numerical studies find that in the lowest LL of a conventional, massive electron system, the CF interactions are sharp and the Fermi surface is stable, while in the first LL a node in the single-particle wavefunction leads to softer CF interactions favorable to pairing.\cite{nayak_non-abelian_2008}
An incompressible quantized Hall state was indeed  observed\cite{willett_observation_1987} in the first LL  of GaAs quantum wells, at filling $\nu=\tfrac52$, though experiments have yet to reveal definitive evidence for nonabelian statistics.

%%%%%%%%%%%%%%%%%% FIGURE 1 %%%%%%%%%%%%%%%%%%%%%%%%%%%
\begin{figure*}
\begin{center}
\includegraphics[width=183mm]{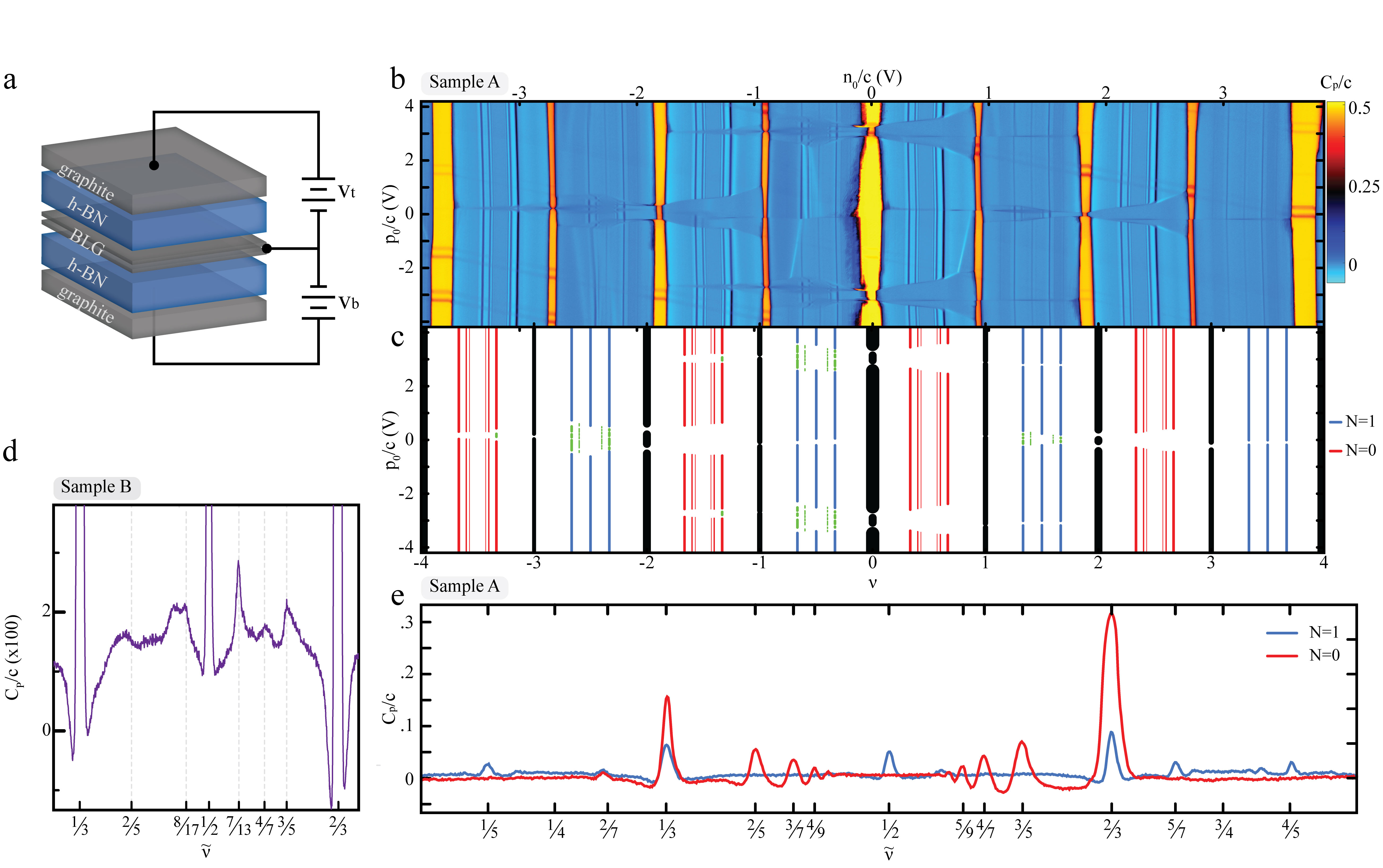} %183
\caption{\textbf{Fractional quantum Hall effect in an all van der Waals heterostructure.}
\textbf{(a)} Device schematic.  A BLG flake is successively encapsulated in both hexagonal boron nitride dielectric and graphite gate layers. Charge density $n$ and layer polarization $p$ are controlled via voltages $n_0/c\equiv(v_t+v_b)$ and $p_0/c\equiv(v_t-v_b)$, where $c$ is the average geometric capacitance of the two gates to the graphene while $v_{t(b)}$ are the applied gate voltages.
\textbf{(b)} Penetration field capacitance $C_P$ at $B=12$~T.  The plot spans the ZLL, showing incompressible quantum Hall states at all integer filling factors, $\nu$, as well as at a multitude of rational $\nu=p/q$.
\textbf{(c)} Orbital character of observed fractional quantum Hall states. As valence electrons fill $N$=0 orbitals (red), we observe odd-denominator fractions consistent with  two-flux composite fermion hierarchy states.  When filling $N$=1 orbitals (blue), only multiples of $\tfrac{1}{3}$ consistently appear from this sequence, with the second most robust state occurring at half-filling. Near orbital and valley level crossings (green), a cascade of interlayer correlated states is observed\cite{halperin_theory_1983}.
\textbf{(d)} Detail of an $N$=1 LL from a second device measured at $B=14\text{T}$ and base temperature, showing particle-hole asymmetric development of incompressible states at $\tilde \nu=3/5$ and $7/13$.
\textbf{(e)} FQH sequences in valence $N$=0 and $N$=1 regions as a function of $\tilde \nu\equiv\nu-\lfloor\nu\rfloor$ for $1<\nu<2$ (blue) and $2<\nu<3$ (red) measured at $p_0/c=-2.0$ and $-2.7$~V, respectively.
The  $N$=0 levels are compressible at half-filling, while the $N$=1 levels show incompressibility peaks.
}
\label{fig1}
\end{center}
\end{figure*}

Bernal bilayer graphene (BLG) is emerging as a new platform for exploring the half-filled LL. Comprised of two aligned graphene layers in direct contact, it has a rich phase diagram that depends on both the electron density $n$ and layer polarization $p$.  A fractional quantum Hall phase was observed\cite{ki_observation_2014} at $\nu=-\tfrac{1}{2}$ in BLG devices suspended in vacuum and gated from below. The interpretation of this state is complicated by the complex structure of the BLG zero energy LL (ZLL), which consists of eight quasi-degenerate components comprising electron spin, a ``valley'' index characteristic of honeycomb systems, and an orbital degeneracy unique to BLG. The spin and valley combine to form an approximately SU(4) isospin, while no such symmetry relates the orbital levels, which are approximately equivalent to the lowest ($N$=0) and first excited ($N$=1) LLs of conventional, massive electrons.
While a nonabelian paired state is expected theoretically when the fractional part of the filling lies in a single $N$=1 orbital, this is difficult to experimentally verify in a singly gated sample.
In devices where the BLG is sandwiched between boron nitride, it can be  gated from both above and below, and the splitting between valley and orbital degrees of freedom can then be controlled using magnetic and electric fields\cite{lee_chemical_2014,maher_tunable_2014}.  A recent experiment\cite{hunt_competing_2016} exploited this control to map out the valley and orbital character of the ZLL, revealing that throughout much of the  accessible parameter space, the valence electrons are fully polarized within a single valley and orbital flavor.  However, even denominator states have not previously been reported in dual-gated devices.

Here we report magnetocapacitance measurements from a new generation of BLG devices, depicted schematically in Fig. 1a.  Unlike previous dual-gated device architectures\cite{lee_chemical_2014,maher_tunable_2014,hunt_competing_2016}, the gate electrodes on both sides of the BLG are made of few-layer graphite flakes, dramatically reducing sample disorder (see Fig. S1).
The sum and difference of the two applied gate voltages, $n_0$ and $p_0$, (see Fig. 1, caption) control the charge density $n$ and layer polarization density $p$ within the bilayer.
Fig. 1b shows the penetration field capacitance, $C_P$, closely related to the thermodynamic compressibility\cite{eisenstein_compressibility_1994}, in a region of the $n_0-p_0$ plane that spans the ZLL, $-4 < \nu < 4$.
Incompressible FQH phases manifest as peaks in $C_P$ locked to the filling factor.
We observe a plethora of new incompressible states at fractional $\nu$ and numerous $p_0$-tuned phase transitions where the state becomes compressible at fixed $\nu$.

We group the  observed FQH sequences into three categories based on the pattern of incompressible phases, indicated by red, blue, and green coloring in Fig. \ref{fig1}c.  In the red regions, we observe sequences of FQH states at valence fillings $\tilde\nu\equiv\nu-\lfloor\nu\rfloor=\tfrac{m}{2 m+1}$. In the blue regions, in contrast, we only observe robust FQH states at $\tilde\nu=\tfrac{1}{3}$, $\tfrac{2}{3}$, and $\tfrac{1}{2}$, with weaker states observed at $\tilde \nu =\tfrac{7}{13}$ and $\tfrac{3}{5}$ (Fig. \ref{fig1}d-e).  The red and blue regions correspond to the experimentally\cite{hunt_competing_2016} determined orbital character ($N$=0 or $N$=1) of the valence electrons, which have different effective interactions.
In red regions, a single $N$=0 component is fractionally filled and the effective interactions are sharp, stabilizing the odd denominator sequence associated with integer quantum Hall states of $2$-flux CFs. We thus ascribe the compressible state at $\tilde \nu=\tfrac{1}{2}$ to the composite Fermi liquid (CFL).
In the blue regions, a single $N$=1 component is fractionally filled and the effective interactions are softer.  This suggests the incompressible state observed at $\tilde\nu=\tfrac{1}{2}$ is a FQH state constructed from paired  CFs. Finally, in the green regions,  $p_0$ induces a level crossing between the eight near-degenerate components\cite{hunt_competing_2016}, and there is a cascade of phase transitions between incompressible states with a structure that depends on the fractional filling, discussed at the end of this letter.

\begin{figure*}[ht!]
\begin{center}
\includegraphics[width=150mm]{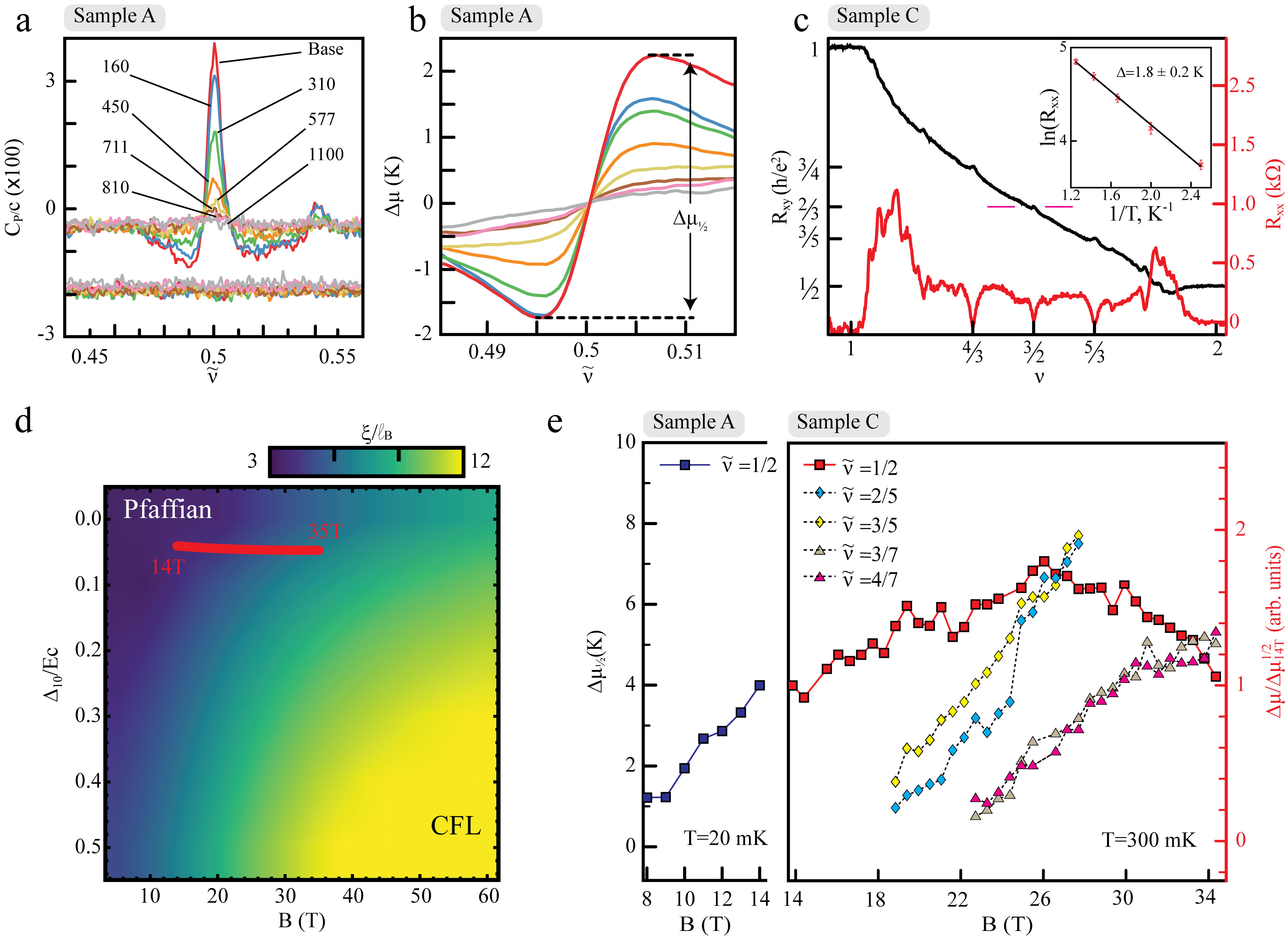}
\caption{
\textbf{The $\tilde\nu = \tfrac{1}{2}$ state.}
\textbf{(a)} Penetration field capacitance (top curves) and dissipation (bottom curves) near $\nu=\tfrac{3}{2}$ at $B=14\text{T}$. Labels denote probe temperature in mK.
\textbf{(b)} Density dependence of the chemical potential, $\Delta\mu\approx \tfrac{e}{k_B}\int(C_P/c)d(n_0/c)$, obtained by integrating curves in (a).
\textbf{(c)} Hall (black) and longitudinal (red) resistance measured in Sample C.  Inset: Arrhenius plot of $R_{xx}\sim e^{-\Delta/(2T)}$ at $\nu=3/2$, from which we obtain $\Delta=1.8\pm.2$K at $B=14$T.
\textbf{(d)} Density matrix renormalization group calculation of the correlation length $\xi$ at $\tilde\nu=\tfrac{1}{2}$ in the $N$=1 level as a function of the energy splitting $\Delta_{10}$ between the $N$=0,1 orbitals in units of Coulomb energy, and the magnetic field  (see Methods). In the lower right corner, the system transitions to the compressible CFL phase. The red line denotes an estimate\cite{hunt_competing_2016} of the trajectory accessed in Fig. 1e.
\textbf{(e)} The thermodynamic gap $\Delta\mu_\frac{1}{2}$ at different $B$ in Sample A  (left panel) and sample C (right panel). Data in the right panel are scaled to the $\Delta \mu_{\frac{1}{2}} (B=14\text{T})$ gap (see Methods).  For energy gaps of other FQH states, see Fig. S7 and Tables S1-S2.
\label{fig2}
}
\end{center}
\end{figure*}

We first quantitatively discuss the observed half-filled FQH states.
In an incompressible FQH state, a finite energy is required to inject an electron or hole. This ``thermodynamic'' gap can be determined\cite{eisenstein_compressibility_1994} from $C_P$, shown in Figure \ref{fig2}a for different temperatures at $B=14 \text{T}$.
We measure this thermodynamic gap by integrating the inverse electronic compressibility ($\partial\mu/\partial n$) with respect to $n$ (Fig. \ref{fig2}b), giving a gap of 4K at the base temperature of our dilution refrigerator (see Methods).
Transport measurements from a second device, meanwhile, show the expected quantized Hall plateau and concomitant longitudinal resistance minimum (Fig. \ref{fig2}c). Temperature dependent transport shows a lower value of the activation gap of  1.8$\pm$.2K at $B=14\text{T}$.   This discrepancy is not surprising\cite{eisenstein_compressibility_1994}.  The  thermodynamic gap measures the energy required to add an entire electron-hole pair, while  thermally activated transport measures the energy cost for injecting a fractionally charged quasiparticle-quasihole pair.
For a half-filled FQH state, the quasiparticle charge is predicted to be $e/4$, in which case the measured activation gap should be roughly a quarter of the thermodynamic gap at T=0\cite{eisenstein_compressibility_1994}.

In a bilayer electron system it is natural to ask whether the incompressible states observed at half-filling are single- or multi-component phases. While the leading theoretical candidates for a single-component even-denominator FQH phase, the paired Pfaffian\cite{moore_nonabelions_1991} and anti-Pfaffian\cite{levin_particle-hole_2007,lee_particle-hole_2007} states, are nonabelian, in multi-component systems the abelian ``331'' phase is more likely\cite{halperin_theory_1983}.
Using  the map of the valence polarization\cite{hunt_competing_2016} (aspects of which were repeated here at higher resolution, see Methods), we find the gapped phase appears in regions where the fractional filling is polarized into a single $N$=1 component.
The situation is thus roughly analogous to the $\nu = \tfrac{5}{2}$ state of GaAs,\cite{willett_observation_1987} where numerics have long predicted a paired phase.  We note, however, that the measured activation gap is several times larger than the largest gaps  measured in GaAs (558 mK\cite{kumar_nonconventional_2010}),  ZnO (90mK\cite{falson_even-denominator_2015}) or suspended BLG (600 mK\cite{ki_observation_2014}).

%%%%%%%%%%%%%%%%%% FIGURE 2 %%%%%%%%%%%%%%%%%%%%%%%%%%%

Despite the superficial similarity, the $N$=1 orbital in BLG differs in two important ways from its counterpart in conventional two-dimensional electron systems in semiconductor quantum wells.
First,  accurately taking into account the four lattice sites in the BLG unit cell breaks the strict equivalence between the $N$=1 LLs of BLG and GaAs, introducing novel tunability.
The $N$=1 LL of BLG includes a combination of the conventional $|0\rangle$ and $|1\rangle$  LL wavefunctions localized on the different sub-lattices of the unit cell (see Methods), with the relative weight of the $|0\rangle$ wavefunction growing with $B$.
The effective interaction depends on the orbital character, so that $B$  continuously tunes the structure of electron-electron interactions within an $N$=1 level.
At low $B$, the wavefunctions are  purely $|1\rangle$-like, with comparatively soft interactions, while at high $B$, they are an equal admixture of $|0\rangle$ and $|1\rangle$ and interactions are consequently sharper. Numerical studies predict that a nonabelian paired phase at lower $B$ should give way to a gapless CFL at sufficiently high magnetic fields\cite{apalkov_stable_2011, papic_topological_2014} (Fig. 2d).
Indeed, we find that the  $\tilde \nu=\tfrac{1}{2}$  gap changes non-monotonically with  $B$ (Fig. \ref{fig2}e), peaking around $B=27$T and then decaying up to the limit of our experiment at $B=35$T.  Over a similar range, we simultaneously observe the emergence of a conventional odd-denominator FQH series typical of the lowest LL, providing further evidence that the effective $N$=1 interactions sharpen with magnetic field (see Methods and Fig. S6). The decrease of the $\tilde \nu=\tfrac{1}{2}$ gap despite an increase in the Coulomb scale $E_C \sim \sqrt{B}$ supports the scenario of a paired-to-CFL transition\cite{metlitski_cooper_2015} at somewhat higher magnetic fields.

Second, particle-hole symmetry is broken differently in BLG compared with GaAs.
Within a single Landau level, the paired Pfaffian and anti-Pfaffian states, which can be understood as  different pairing channels, are degenerate due to a particle-hole symmetry (effected by the transformation $\tilde{\nu} \leftrightarrow 1 - \tilde{\nu}$).
Including scattering between LLs  breaks this symmetry and determines the ground state.
While the subject of longstanding debate, recent numerical agreement between exact diagonalization and DMRG methods suggests that the $\nu=\tfrac{5}{2}$ state of GaAs is in the anti-Pfaffian phase \cite{rezayi_breaking_2011, zaletel_infinite_2015,rezayi_landau-level-mixing_2017}.
However, LL scattering is dramatically different in BLG: scattering between the ZLL and the $|N|\geq 2$ levels only breaks particle-hole symmetry weakly, while scattering within the ZLL breaks it strongly due to the small splitting ($\Delta_{10}\approx .1 E_C$) between $N$=0 and $N$=1 levels (see Methods).  In our experiment, particle-hole symmetry breaking manifests in the fractions observed in the $N$=1 LL. We find incompressible states at $\tilde{\nu} = \tfrac{7}{13}$ and $\tfrac{3}{5}$ (Fig. 1d), the particle-hole conjugates of what is observed in GaAs where unconventional states were observed at $\tfrac{6}{13}$ and $\tfrac{2}{5}$.\cite{kumar_nonconventional_2010}
To address these differences, we perform comprehensive DMRG calculations which account for the $B$-dependent mixed orbital character and screening from filled $|N|\geq 2$ LLs, while non-perturbatively accounting for scattering between the $N$=0 and 1 orbitals of the ZLL (Fig. \ref{fig2}, see Methods for computational details).
We find that, in contrast to GaAs \cite{rezayi_breaking_2011, zaletel_infinite_2015,rezayi_landau-level-mixing_2017}, the Pfaffian phase is strongly preferred over the anti-Pfaffian over the experimentally accessible range.
Suggestively, $\tfrac{7}{13}$ (as well as $\tfrac{8}{17}$, where a weaker feature is also observed) is the predicted filling of the first ``daughter'' state of the Pfaffian phase\cite{levin_collective_2009}.

%%%%%%%%%%%%%%%%%% FIGURE 3 %%%%%%%%%%%%%%%%%%%%%%%%%%%
\begin{figure*}[ht!]
\begin{center}
\includegraphics[width=183mm]{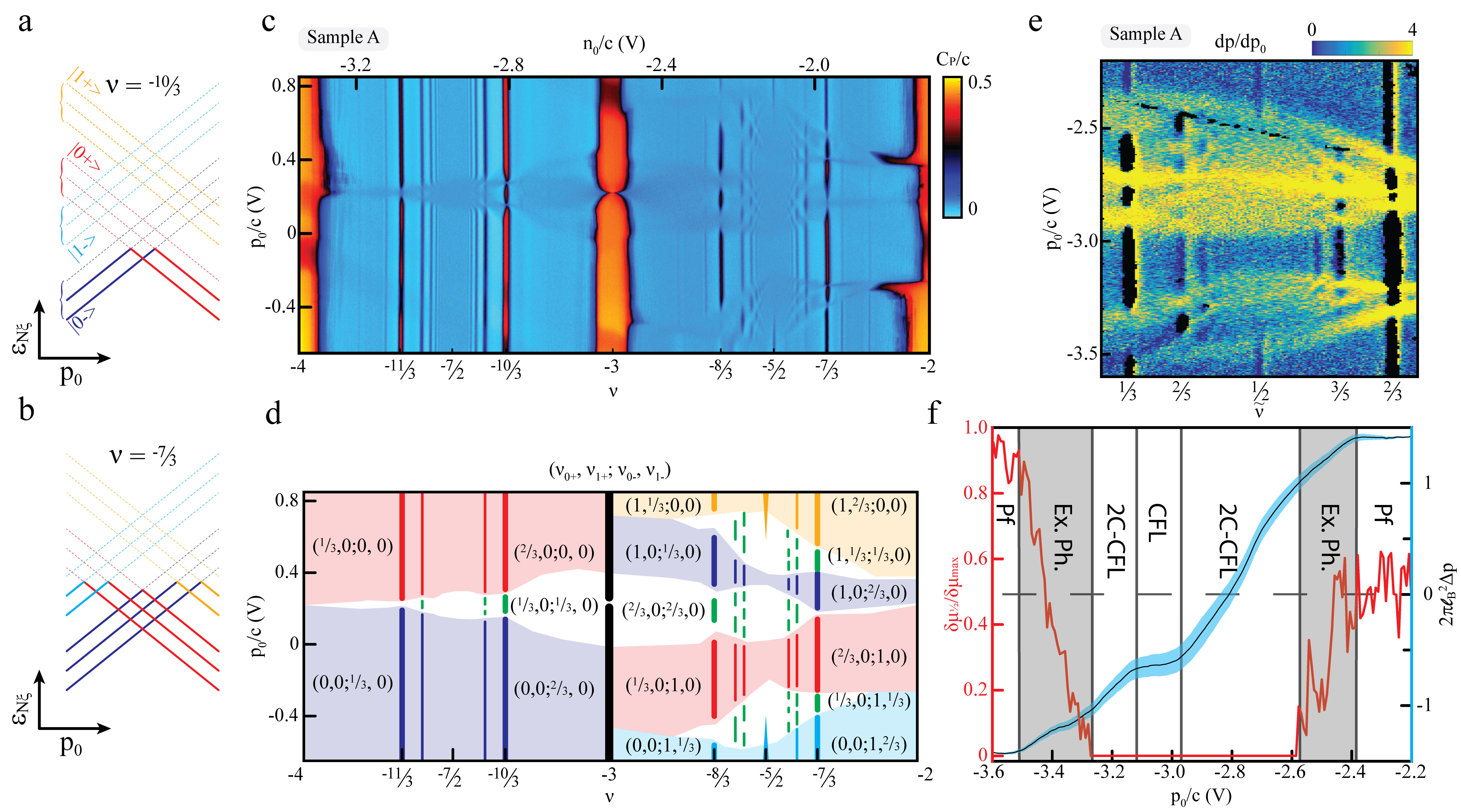} %183
\caption{
\textbf{Interlayer correlated FQH states}.
Single particle energy level ($\epsilon_{N\xi}$) crossing and level filling diagram as a function of $p_0$, for \textbf{(a)} $\nu=-10/3=-4+2/3$ and \textbf{(b)} $\nu=-7/3=-4+5/3$.  Occupation of the levels in increments of $\nu=1/3$ is represented by schematically showing each LL as divided into three branches. Three distinct phases are expected by filling the two lowest lying `branches' in (a). In (b), crossings now involve both $N$=0 and $N$=1 levels, and 6 distinct phases are expected.
\textbf{(c)} Measured $C_P$ for $-4<\nu<-2$ near $p_0/c=0$ at $B=12$T.
\textbf{(d)} Annotated phase diagram for the range depicted in (c). Occupations of the four relevant orbitals are indicated for each fractional multiple of $1/3$.  Shaded areas correspond to regions where the fractional filling lies entirely within one orbital.  Coloring follows the scheme in (a-b).
\textbf{(e)}
Layer polarizability $\tfrac{dp}{dp_0}$, measured over an analogous quadruple level crossing at high negative $p_0$ for $-1<\nu<0$.
\textbf{(f)} Black curve shows the integrated change in polarization, $2\pi\ell_B^2\Delta p=2\pi\ell_B^2\int \tfrac{\partial p}{\partial p_0}dp_0$, measured in the regions immediately adjacent to $\tilde\nu=\tfrac{1}{2}$, with shading indicating $1\sigma$ confidence interval (see Methods).
Red curve shows the $\tilde \nu=\tfrac{1}{2}$ charge gap. Vertical lines demarcate  distinct regimes distinguished  by their compressibility and polarizability.  The incompressible and unpolarizable regions are the Pfaffian phase; incompressible but polarizable  regions are the excitonic phase discussed in the main text; compressible but polarizable regions  are presumably two-component CFLs, and the compressible and unpolarizable is a one component CFL.
exciton condensate.  Further depolarization collapses this phase to the 2C-CFL, and eventually the CFL.
\label{fig3}}
	\end{center}
\end{figure*}

Our results suggest encapsulated BLG has certain advantages over GaAs as a platform for interferometric detection of nonabelian quasiparticles\cite{nayak_non-abelian_2008}.
First, the large energy gap and small correlation length relative to GaAs may reduce bulk-edge coupling that is detrimental to interferometric probes\cite{von_keyserlingk_enhanced_2015}, while exponentially suppressing the density of thermally-activated quasiparticles.
Second, hBN gate dielectrics can be made almost arbitrarily thin, allowing one to engineer edges and quantum point contacts using sharp electrostatic potentials. Recent experiments have demonstrated long coherence lengths in the quantum Hall regime along such gate-defined edges\cite{wei_mach-zehnder_2017}.
Finally, the putative Pfaffian state at $\nu = -\tfrac{1}{2}$ in BLG would have fewer edge modes than the anti-Pfaffian state at $\nu = \tfrac{5}{2}$ in GaAs, making the former a preferable candidate for interferometry.
Even without phase coherent transport measurements, the thermodynamic measurements presented here, carried to lower temperatures, can be used to probe topological ground state degeneracy\cite{cooper_observable_2009},  providing smoking-gun evidence for nonabelian statistics in the near future.

In addition to control over the total density $n$ and the effective interactions, the dual-gated architecture allows us to tune level crossings between the eight components of the ZLL.  Within the ZLL, the two valleys are supported on opposite layers, so the electric field ($p_0$) acts like a ``valley Zeeman'' field and the layer polarization ($p$) can be used to infer valley polarization.
A schematic of the single particle energies near $p_0 \sim 0$ is shown in Figs. \ref{fig3}a-b.  Four single particle levels are involved in the crossing, which we label by their orbital ($N$ = 0, 1) and valley ($\xi = \pm$)  indices (we suppress the spin here, since tilt $B$-field measurements show that the spin-polarization is unchanged across the transition).
Because the two valleys are distinguished by their crystal momentum,  the  tunneling between them vanishes in the absence of short-range disorder and the crossing between the  levels is unavoided, as supported by the sharp transition at $\nu = -3, p_0 \sim 0$ (Fig.~\ref{fig3}c).
Hence, unlike the dependence on $B$, the $p_0$-dependence across the transition cannot be understood as a continuous tuning of the interaction potential.
Indeed, when charge is separately conserved in each valley, the inter-valley polarization cannot change continuously without closing the neutral gap - just as the charge gap vanishes in a compressible system, the neutral gap vanishes in a polarizable system. During such depolarization, the charge gap may or may not close.

In Fig \ref{fig3}c, we show a detailed plot of $C_P$ for $p_0 \sim 0$.
Focusing first on $-4 < \nu < -3$, we see that for the best developed odd-denominator $\tilde{\nu} = \tfrac{m}{2 m + 1}$ states, $|m|+1$ distinct high-$C_P$ incompressible regions are visible, separated by $|m|$ low-$C_P$ transitions.
Referring to Fig. \ref{fig3}a, the crossing is predicted to transfer valence filling $\tfrac{m}{2 m + 1}$ between the $N$=0 orbitals of opposite valleys, so the observed $m$ compressibility spikes presumably occur when filling $\tfrac{1}{2 m + 1}$ transfers between the two valleys. This scenario is in  agreement with expectations from composite Fermion theory, which predicts two-component correlated states\cite{halperin_theory_1983,jain_composite-fermion_1989} at fillings $(\nu_+, \nu_-) = \tfrac{(m_+, m_-)}{2 (m_+ + m_-) + 1}$, (here $\nu_{\pm}$ is the filling of valley $\xi=\pm$) separated by phase transitions where the gap closes. The state at $\nu = -4 + \tfrac{2}{3}, p_0 = 0$, for instance, corresponds to $m_{\pm} = -1$ and we ascribe it to a previously unobserved valley SU(2) singlet.
For $-3 < \nu < -2$, states at filling $\nu = -4 + \tfrac{3 m + 1}{2m + 1}$  show $3 m + 1$  transitions.  Four levels are involved in these transitions.  At high $p_0$, one $N$=0 level is completely filled and the fractional filling resides in the $N$=1 level of the same valley.  As $p_0$ is decreased, occupation is transferred according to the levels shown in Fig. \ref{fig3}b, consistent with the observed strengths of the gapped phases, whose $C_P$-dips are strongest when only $N$=0 orbitals are involved.

For odd denominator states, the high compressibility observed when the system changes polarization indicates that the gap for charged excitations also closes. This is not always the case at  $\tilde{\nu} = \tfrac{1}{2}$, where the charge gap in the single component $N$=1 regimes at large and small $p_0$  fades  gradually into the level crossing.  We can quantify this transition by directly measuring the layer polarization (see Methods). Fig. \ref{fig3}e shows the layer polarizability ($\partial p/\partial p_0$) over a similar region of four-level crossings. In contrast to the odd denominator fractions, where the spikes in polarizability are concentrated on the spikes in compressibility, near $\tilde \nu=\tfrac{1}{2}$ there is a region of $p_0$ where the polarization changes only gradually while the charge gap remains finite.
Fig. \ref{fig3}f shows the measured charge gap alongside the integrated change in layer polarization across the level crossing.
We find that the charge gap persists across a wide range of valley valence fillings between $(\nu_+, \nu_-) = (1.5, 0)$ and $(1.33, .17)$.

The observation of polarizability coexisting with incompressibility has intriguing implications. At least in the clean limit where charge is conserved separately in each valley (a limit supported by the sharp transition at $\nu = -3, p_0\sim0$), finite polarizability requires a vanishing neutral gap, implying the existence of a new phase: a gapless fractionalized insulator.
Microscopically, because the layers are atomically close, the finite polarization presumably arises from a finite density of inter-valley (e.g., inter-layer) excitons, and the finite polarizability implies these neutral excitons are gapless.
This is reminiscent of quantum Hall bilayers at $\nu=1$, where a charge gap also coexists with a vanishing neutral gap.  The transition is thus distinct in  microscopic character from the Pfaffian-to-CFL transition predicted at high B in a single-component level (Fig. 2e), where the charge and neutral gap would vanish in tandem.

Theoretically, the accompanying fractionalization at $\tilde \nu=\tfrac{1}{2}$ leaves several possibilities for the ultimate collective ground state---and indeed even the quantum statistics---of  inter-valley excitons\cite{barkeshli_fractionalized_2016}.
Most simply, the incompressible exciton phase could be disorder dominated: as charge is transferred between valleys, the resulting density of excitons is trapped by local potential variations in a mechanism similar to that which stabilizes FQH plateaus over a finite range of $\nu$. However, as is evident in Figs. 2b-c, the even denominator state itself is only stable to pure charge doping up to  $\Delta\nu\approx.005$, more than an order of magnitude less than the occupation change ($\Delta\nu_+\approx .17$) of the $N$=1 orbital implied by the depolarization measurement. Absent this mechanism, the incompressible exciton phase may host such phenomena such as interlayer phase coherence or an emergent Fermi surface, which can be distinguished experimentally by probing thermal transport or interlayer Coulomb drag.

\section*{Contributions}
AAZ, EMS, and HZ fabricated  devices A, B and C, respectively. TT and KW synthesized the hexagonal boron nitride crystals. AFY and CK built the measurement electronics. AAZ, HZ, EMS and AFY acquired and analyzed the experimental data. MPZ performed the DMRG calculations.  AAZ, MPZ, AFY wrote the paper.

\section*{Acknowledgments}
The authors acknowledge experimental assistance of Brunel Odegard and Joshua Island, and discussions with Maissam Barkeshli, Cory Dean, Eun-Ah Kim, Roger Mong, Chetan Nayak,  Zlatko Papic, Steve Simon and Ady Stern. Magnetocapacitance measurements were funded by the NSF under DMR-1636607. A portion of the nanofabrication and the transport measurements were funded by ARO under proposal 69188PHH. AFY acknowledges the support of the David and Lucile Packard Foundation.  Measurements above 14T were performed at the National High Magnetic Field Laboratory, which is supported by National Science Foundation Cooperative Agreement No. DMR-1157490 and the State of Florida. The numerical simulations were performed on computational resources supported by the Princeton Institute for Computational Science and Engineering (PICSciE). EMS acknowledges the support of the Elings Fellowship.
 \setcounter{figure}{0}
 \setcounter{section}{1}
 \setcounter{subsection}{0}

\section{Methods}
\renewcommand{\thefigure}{M\arabic{figure}}
\renewcommand{\thetable}{M\arabic{figure}}
\renewcommand{\theequation}{M\arabic{equation}}

\subsection{Device fabrication}
Devices were assembled using a dry transfer method based on van der Waals adhesion\cite{wang_one-dimensional_2013}, with top and bottom gates as well as the electrical contacts to the hBN-encapsulated Bernal stacked BLG device made from $\sim10$~nm-thick graphite flakes.  As shown in Fig. \ref{fig:comparison}, this leads to a significant improvement in sample quality as compared to samples with evaporated metal gates. Care was taken in order to match the top and bottom hBN thicknesses, which were approximately 40-50 nm for all three devices.  The resulting mismatch in geometric capacitance was determined to be $\delta \equiv (c_t - c_b)/(c_t+c_b)=0.018$ for Device A, and was $<.05$ for all devices.  Despite efforts to rotationally misalign the hBN crystals proximate to the graphene bilayer, the effects of an intermediate wavelength moire pattern\cite{dean_hofstadters_2013,ponomarenko_cloning_2013,hunt_massive_2013} are visible on one of the layers at high density in device A (Fig. \ref{moire}).  The resulting staggered sublattice potential is responsible for the asymmetry upon inversion of $p_0$ in some of the transitions visible in Fig. 1 of the main text.  While no secondary fans were observed in  devices B and C, similar offsets (albeit of differing magnitudes and signs) were observed in these devices as well, indicating that coupling to one or more of the hBN layers remains relevant (See Figs S3, S4, and S5).
Devices A and C were patterned by using a dry ICP etch in a mixture of CHF$_3$+O$_2$, and the graphite contacts were themselves contacted along the edge with a Cr/Pd/Au metal stack.  Device B was first exposed to a XeF$_2$ atmosphere to remove a sacrificial top hBN layer used for stack assembly, annealed at 400~$^{\circ}$C in forming gas, and the graphite contacts and gates were area contacted with a Ti/Au metal stack.

\subsection*{Capacitance circuit}

\begin{figure*}[ht]
\begin{center}
\includegraphics[width=183mm]{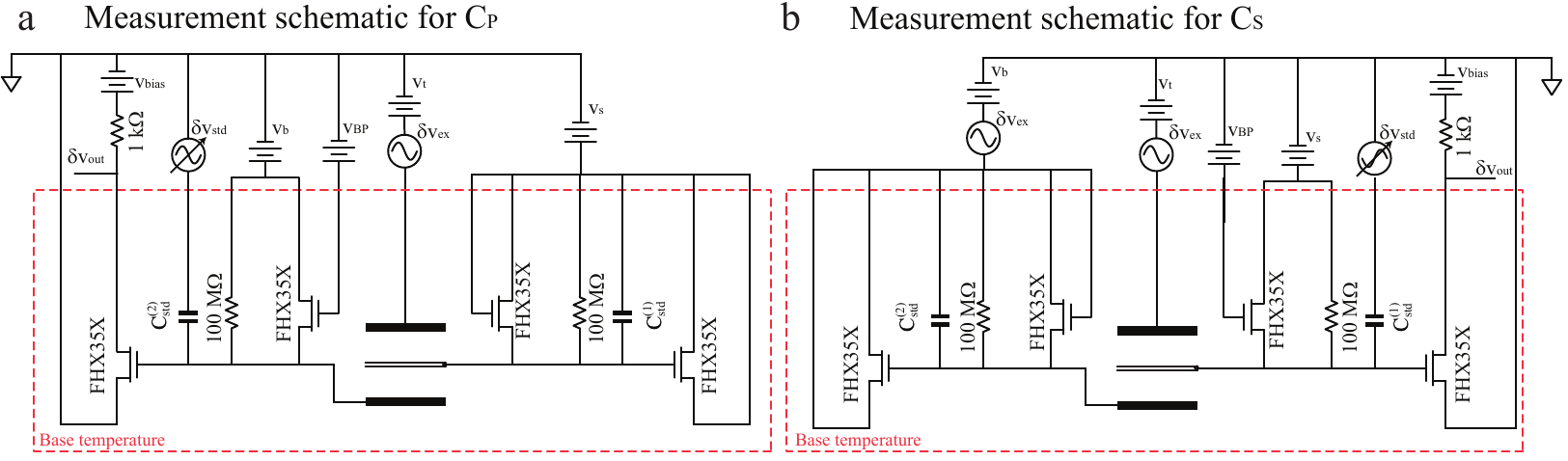} %183
\caption{Electrical schematics for the penetration field capacitance measurement and symmetric capacitance measurements.}
\label{schematic}
	\end{center}
\end{figure*}

To measure the small capacitances of our device with high sensitivity, we utilize a cryogenic impedance transformer based on an FHX35X high electron mobility transistor\cite{ashoori_single-electron_1992} in a bridge configuration. This measurement works by effectively disconnecting the sample capacitance from the large capacitance of the cryostat cabling: the BLG device is connected only to the gate of the HEMT (with input capacitance of a few hundred fF) and a large resistor (100 MOhms), with additional $\sim$~1pF of stray capacitance. While the HEMT is operated at or below unity voltage gain to minimize power dissipation, it effectively transforms the picofarad sample impedance into a 1k$\Omega$ output impedance.  At 50 kHz, and assuming cryostat cable capacitances of $\approx$1nF, this translates to a power gain of approximated 1000.

As described in detail in references \onlinecite{young_capacitance_2011} and \onlinecite{hunt_competing_2016}, measurements of three distinct capacitances---the top gate, bottom gate, and penetration field capacitance---provide a complete reconstruction of both the charge and layer polarization in the bilayer as a function of applied voltages.  For devices in which top and bottom gate geometric capacitances are symmetric ($\delta\equiv\frac{c_t-c_b}{c_t+c_b}\rightarrow 0$, see supplementary information of \onlinecite{hunt_competing_2016} for a detailed derivation), one accesses the derivatives of the total density ($n$) and layer polarization ($p$) with respect to the applied voltages $n_0$ and $p_0$ through the relations
\begin{align}
\frac{\partial n}{\partial n_0} &\approx \frac{C_S}{c}     \label{eqdndn0}\\
\frac{\partial n}{\partial p_0} &\approx \frac{C_A}{c}     \\
\frac{\partial p}{\partial p_0} &\approx \frac{2c_0}{c}\frac{C_S+4C_P+2c}{c} \label{eqdpdp0}\\
\frac{\partial p}{\partial n_0} &\approx \frac{2c_0}{c}\frac{C_A}{c} ,
\end{align}
where $c=(c_t+c_b)/2$ is the average of the top and bottom gate geometric capacitances, $c_0\gg c$ is the interlayer capacitance of the bilayer itself, and $C_{S(A)}=C_T\pm C_B$ are the symmetric and antisymmetric measured gate capacitances per unit area.  The approximation is well justified in our devices, where $\delta=.018$.

In order to measure all three quantities without warming up the sample, low-impedance access to all three terminals of the device must be possible in situ, in apparent conflict with the desire to maintain the amplifier input at high impedance.   This problem is solved using a cryogenic multiplexer constructed out of two additional HEMTs, which allows either the bottom gate or the bilayer graphene flake to be brought to a high impedance.  Figure \ref{schematic}a shows a simplified electrical schematic  for measuring $C_P\equiv\partial n_T/\partial v_b$, where $n_T$ is the  charge  on the top gate while $v_b$ is the bottom gate voltage. $v_b\approx-.3\text{V}$ is fixed to set the transistor operating point, $v_{bias}\approx 25 \text{mV}$ is chosen so that no heating of the probe is observed (see the next subsection on likely electron temperature).  Additional DC voltages on the top gate ($v_t$) and applied to the ohmic contacts of the graphene ($v_s$) are varied to control $n_0$ and $p_0$, defined as
\begin{align}
n_0&=c_t(v_t-v_s)+c_b(v_b-v_s)\approx c(v_t+v_b-2v_s)\\
p_0&=c_t(v_t-v_s)-c_b(v_b-v_s)\approx c(v_t-v_b).
\end{align}

To measure the differential capacitance, a fixed AC excitation  is applied to top gate ($\delta v_{ex}$). A second AC excitation at the same frequency is applied to a standard capacitor $\delta v_{std}$, with phase and amplitude chosen to balance the bridge ($\delta v_{out}=0$). Crucially, for the $C_P$ measurement the `bypass' HEMT attached to the bottom gate is driven deep into depletion ($v_{BP}\approx v_b-.6V$), maintaining the high impedance of the measurement transistor input.  A final DC voltage is applied to all pins of the second amplifier ($v_s$). Because the FHX35X is in depletion mode, in this configuration it shorts out the 100 M$\Omega$ resistor, providing a low impedance connection to the sample. To measure $C_S$, the situation is reversed (Figure \ref{schematic}b), with the transistor amplifier on the sample now active and the bypass transistor on the bottom gate short circuited, providing a low impedance path for the bottom gate voltage excitation. Although not focus of the current work, $C_A$ can also be measured by applying opposite phase excitation signals to the top and bottom gate\cite{hunt_competing_2016}.
The frequency and excitation voltages for data shown in all figures are shown in Table \ref{freq}.
\begin{table}[ht]
\caption{Frequency and excitation voltages used in capacitance data presented in main text, methods, and supplementary  figures.}
\label{freq}
\begin{tabular}{|c|c|c|}
\hline
Figure&$\delta V_{ex}$ (mV) & $f$ (kHz)\\
\hline
1b		&1.4	&	71.77	\\\hline %d111
1d		&0.22	&	85.77	\\\hline %d64
1e		&0.9	&	2.48	\\\hline %d1269, d1270
2a,b	&0.9	&	2.48 	\\\hline %d1287-d1305
2e (L)	&0.9	&	2.48	\\\hline %d1287-d1305
2e (R)	&1.6		&17.7			\\\hline %d267
3c		&0.9	&71.77		\\\hline %d342
3e		&2.96	&8.17		\\\hline
M2a-c   &2.96   &8.17       \\\hline %d1378, d1380,
M2d		&0.9	&8.17		\\\hline %d1355
%S1(L)   & "Andrea's Notebook"      &           \\\hline %d488
S1(R)   &2.96      &71.77           \\\hline
S2		&2.1	&81.7		\\\hline %d813
S3		&1		&71.77		\\\hline
S4		&.2		&87.77		\\\hline
S5		&4.7	&84.77		\\\hline
S6      &1.6       &17.7           \\\hline %d267
S7      &0.9       & 2.48          \\\hline %d1287-d1305
\hline
\end{tabular}
\end{table}

All measurements below 14T were performed in a dry dilution refrigerator with a base temperature of $\approx$ 10 mK. However, the sample temperature was likely higher due to heating from the cryogenic amplifiers, which were directly connected to the sample and only a few millimeters away.  While we do not have a  thermometer for the electron temperature in our devices, recent tunneling experiments, which use an identical amplification scheme in a dilution fridge with a similar base temperature, do allow for in situ thermometry\cite{jang_sharp_2016}.  In these experiments, electron temperatures below 100 mK are possible only with careful thermal isolation and heat sinking of the HEMT, which is a heat source directly tied to the sample\cite{Jang_none_2017}. Because we have not taken these precautions in our setup, our electron temperature is likely no less than 100mK for the capacitance measurements.  Indeed, data taken at the NHMFL in Tallahassee in a 3He system with a base temperature of 300mK do not look qualitatively different.

\subsection*{Measurement of electronic compressibility and thermodynamic energy gaps}
Most of the experimental data presented in the main text are $C_P$, which we interpret as proportional to electronic compressibility.  As is clear from Eqs. \ref{eqdndn0}-\ref{eqdpdp0}, this is not generally true: solving those equations for $C_P$ gives
\begin{align}
C_P&=\frac{c}{2}-\frac{1}{4}\frac{\partial n}{\partial n_0}+\frac{c}{8c_0}\frac{\partial p}{\partial p_0}.
\end{align}
The third term in this equation is unique to multilayers, denoting the layer polarizability.  Near layer polarization phase transitions, where the layer occupation changes rapidly over a small range of $p_0$, this term can be large; however, in the single component regimes where, for example, we perform our measurements of the thermodynamic gaps, this term is tiny.  In these regimes, charge cannot move easily between the layers, and the bilayer system behaves as a single layer system (measurements of $\partial p/\partial p_0$, discussed below, are featureless in this regime).  In this case, the conventional quantum capacitance\cite{luryi_quantum_1988} formula applies, so that
\begin{align}
C_P&=\frac{c}{2}-\frac{1}{4}\left(\frac{1}{2c}+\frac{1}{\partial n/\partial \mu}\right)^{-1}\\
&=\frac{c^2}{2c+\partial n/\partial \mu}\\
\end{align}

\begin{figure*}[ht]
	\begin{center}
\includegraphics[width=183mm]{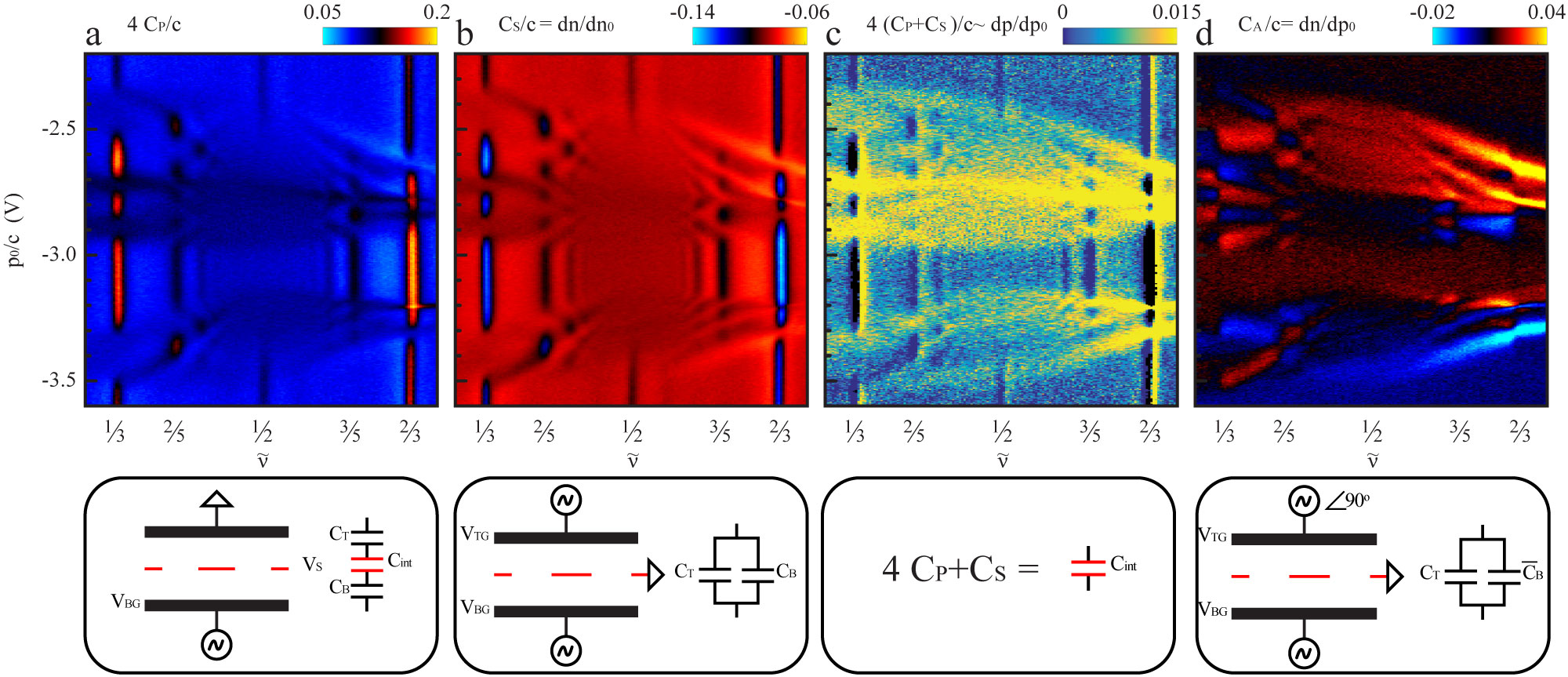} %183
\caption{Thermodynamic properties of interlayer correlated states. \textbf{(a)} Penetration field capacitance $4\times C_\text{P}/c$, \textbf{(b)} Symmetric capacitance $C_\text{S}/c$, \textbf{(c)} $4C_\text{P} + C_\text{S}$, proportional to the polarizability, \textbf{d)} anti-symmetric capacitance $C_\text{A}$ . As presented, $C_\text{P}/c$ and $C_\text{S}/c$ were offset by constant parasitic capacitance levels $C_\text{par}/c=-0.55$ and $C_\text{par}/c=-2.84$ respectively.}
\label{s15}
	\end{center}
\end{figure*}

For the fractional quantum Hall states under discussion, the modulation of $C_P$ over the background value is small ($C_P\ll c/2$), implying $\partial n/\partial \mu\gg c/2$  and $n\approx n_0$.  Thus the change in chemical potential follows as
\begin{align}
\Delta\mu_{12}
&=\int_{n^{(1)}}^{n^{(2)}} \frac{\partial \mu}{\partial n}dn\\
&=\int_{n^{(1)}}^{n^{(2)}}\frac{C_P/2}{c/2-C_P}d\left(n/c\right)\\
&\approx \int_{n_0^{(1)}}^{n_0^{(2)}} \frac{C_P}{c}d\left(n_0/c\right)\label{eqgap}
\end{align}
We take the gap as the difference between the maximum and minimum in $\Delta \mu$ near a FQH state, as shown in Fig. 2b of the main text.  As can be seen from Eq. \ref{eqgap}, quantitative measurement of $\Delta\mu$ requires accurate knowldge of $c$.  This is calibrated in situ from the capacitance in an integer quantum Hall gap (where the full c/2 penetration capacitance obtains).  Unfortunately, this data was not be acquired in the limited time available at the National high magnetic field lab.  For this reason, the gaps in Fig. 1e (right panel) are presented only in relative units, normalized to their value at B=14T.

Interpretation of any measured capacitance as a thermodynamic derivative requires that the sample is sufficiently conductive to fully charge over a time scale comparable to the inverse measurement frequency\cite{goodall_capacitance_1985}. At low temperature and high magnetic fields, our sample becomes strongly insulating at all integer and many fractional filling factors, precluding this condition being satisfied at high frequency. This is unimportant for identifying the existence of fractional quantum Hall phases, which manifest as $C_P$ peaks regardless of whether the contrast mechanism is due to low electronic compressibility or low bulk conductivity; however, it is critical to be in the low frequency limit for any quantitative analysis of energy gaps following Eq. \ref{eqgap}.  Failure to charge manifests as an increase in the out of phase, dissipative signal in the complex capacitance, $\tilde C=C+iD$, where we have plotted `C' throughout the text.  We can monitor charging across the parameter range by plotting `D'.  In order to measure energy gaps, we decrease the frequency until no features are visible in D.  In this limit, it is justified to integrate $C_P$ to extract energy gaps.

\subsection*{Measurement of layer polarizability, $\partial p/\partial p_0$}

To measure $\partial p/\partial p_0$, as shown in Fig. 3e of the main text, we measure penetration field ($C_P$) and symmetric capacitance ($C_S$) over a range of $n_0$ and $p_0$ corresponding to the four-level crossing described in the main text Figs. 3a-b.  As can bee seen in Fig. \ref{fig:comparison}d, there are diagonal features in $C_S$ within the  $n_0-p_0$ plane that correspond to constant gate voltage.  We ascribe these features either to single-gated regions or to the appearance of incompressible LL gaps in the graphite gates themselves. To avoid them, we perform the measurement in the nearly identical level crossing at $p_0/c\approx -3V$ for $-1<\nu<0$, where diagonal features are not observed in the regime of the phase transition.
$C_P$ and $C_S$ in this regime are shown in Fig \ref{s15}, along with their weighted sum, $4C_P+C_S$ as per Eq. \ref{eqdpdp0}.  The two measurements were calibrated against each other by measuring $C_T$, the top gate capacitance, accessible in either configuration.

FQH gaps show strong features in the $\partial p/\partial p_0$ data.  This is spurious: digital subtraction of the two data sets measured with different amplifiers leads to systematic errors where $C_P$ and $C_S$ have large gradients, for example in FQH gaps.  There, small offsets in $n_0$ and  nonlinearities in the amplifier chain can lead to incomplete cancellation of the two signals, leading to the visibility of the gaps themselves in Fig. 3e.
This is confirmed by the measurement of $\frac{dp}{dn_0}$: if the changes in polarization were real, they would manifest in this quantity as well.  $\frac{\partial p}{\partial n_0}$ is shown in Fig. \ref{s15}d. Because $\frac{\partial p}{\partial n_0}$ can be measured directly with a single amplifier(see, for example, \cite{hunt_competing_2016}), subtraction-induced systematic errors are automatically canceled.  We find no change in polarization of the $\tilde \nu=\frac{1}{2}$ state relative to its immediate background, indicating that the features in $\partial p/\partial p_0$ at fractional filling are indeed spurious.  We thus integrate $\frac{\partial p}{\partial p_0}$ at $\tilde{\nu}=0.495$ and $\tilde{\nu}=0.505$ and average the results to determine .  Figure \ref{s15}c shows raw $\frac{\partial p}{\partial p_0}$ data,

To determine the layer polarization change for arbitrary $\nu$ and $p_0$, we integrate the measured $C_S+4C_P$ signal, as

\begin{align}
\Delta p_{12}&=2\pi\ell_B^2\int_{p_0^{(1)}}^{p_0^{(2)}} \frac{\partial p}{\partial p_0} dp_0\\
&=2\pi\ell_B^2\frac{2c_0}{c} \int_{p_0^{(1)}}^{p_0^{(2)}} \left(\frac{C_S+4C_P+2c}{c}\right) dp_0\\
&=2\pi\ell_B^2 2c_0 \int_{p_0^{(1)}}^{p_0^{(2)}} \left(\frac{C_S+4C_P+2c}{c}\right)  d(p_0/c) \label{Aimplicit}
\end{align}
where $\Delta p$ is expressed in filling factor units and all capacitances are understood to be in units of particle number per area per volt.

The measured data may be subject to an arbitrary offset due to differing parasitic capacitances in the $C_S$ and $C_P$ measurements which need not cancel.  In addition, an overall magnitude error can lead to a systematic under- or over estimate of the change in polarization.  To compensate these errors, we take advantage of the fact that we know, from band structure, the total layer polarization change that must occur between the two extremes in Fig. 3e (and Fig. \ref{s15}a-d). At the bottom of these plots, an N=0 orbital on the bottom layer is fully occupied, while an $N=1$ orbital on the bottom layer has occupation $\tilde \nu$.  At the top, these orbitals are empty while an N=0 orbital on the top layer is fully occupied while the $N=1$ orbital on the top layer has occupation $\tilde \nu$.  Thus, for a given $\tilde \nu$ in this regime, we expect a total occupation transfer of $\Delta\nu=2(1+\alpha \nu)$, where $\alpha=.84$ is the layer polarization of the $N=1$ orbitals at 12T determined from tight binding calculations.  We use this to fit two constants to the integrated data, $a$ and $b$, so that
\begin{equation}
a\int_{-3.6}^{-2.2}\left(\frac{C_S+4C_P}{c}-b\right)d\left(\frac{p_0}{c}\right)=2(1+\alpha\nu)
\label{AB}\end{equation}
where the contour of integration is at constant $\tilde \nu$ from the bottom to the top of Fig. \ref{s15}c.
The constant $b$ can be determined directly by measuring the background level in a single component region, where we expect $\partial p/\partial p_0=0$, and this background is subtracted from the $4C_P+C_S$ data set shown in Fig. \ref{s15}c. The error with which this background can be determined is $\sigma_b=2.3\times 10^{-4}$.
We then calculate $a$ from Eq. \ref{AB} by performing this integration for 35 combinations of 7 different regions of $\tilde \nu$, from which we find $a=(422\pm 22)$~V$^{-1}$.  Following Eq. \ref{Aimplicit} and taking
$d_{BLG}=.335\mathrm{nm}$ and $B=12\mathrm{T}$
allows us to extract the interlayer dielectric constant of the graphene bilayer, $\epsilon_{BLG}$ since,
  \begin{align}
a&=4\pi c_0\ell_B^2\\
a&=4\pi \frac{\epsilon_{{\mathrm{BLG}}}}{d_\mathrm{BLG}}\frac{\epsilon_0}{e}\ell_B^2\\
\epsilon_{\mathrm{BLG}}&=\frac{a}{113.6\mathrm{V^{-1}}}=3.71\pm.19
\end{align}
roughly consistent (to within 25\%) with $\epsilon_{BLG}=2.8$ determined in recent experiments at higher magnetic fields\cite{hunt_competing_2016}.  We note that $\epsilon_{BLG}$ receives contributions from filled Landau levels, and may well be field dependent.  To generate Fig. 3f of the main text, we integrate along $p_0/c$, for example
\begin{align}
\Delta p(p_0)&=a\int_{-3.6}^{p_0/c}\left(\frac{C_S+4C_P}{c}-b\right)d\left(\frac{p_0}{c}\right)\\
\sigma_{\Delta p}(p_0)&=\frac{\sigma_a}{a}\Delta_p(p_0)+\sigma_b(p_0/c+3.6)
\end{align}
Because the layer polarization is known also for $p_0/c=-2.2$~V, either the increasing or decreasing integral are equivalent, and error bars are defined by the lesser of the two. The largest uncertainty thus obtains approxiately midway between the extremes near $p_0/c=- 2.9$ V.

\begin{figure}[ht]
	\begin{center}
\includegraphics[width=.8\columnwidth]{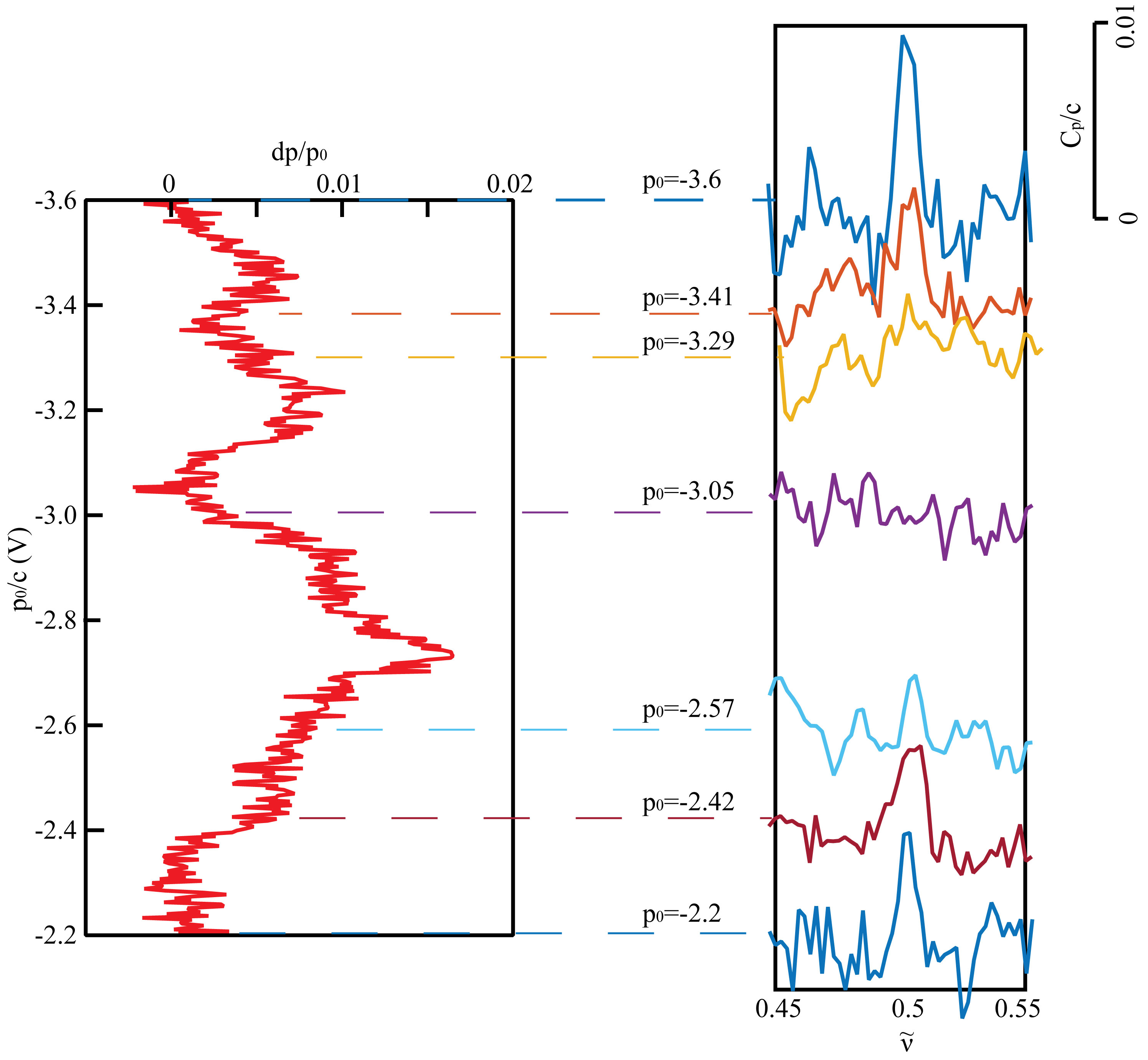} %183
\caption{$C_\text{P}+4C_\text{S}$ near $\tilde{\nu}=1/2$ (left panel) and $C_P$ across the $\tilde{\nu}=1/2$ state (right panel), showing the persistence of the gap with layer polarization. The gap feature fades with increasing $p$ but does not completely vanish until $p_0/c\approx -3.3$ (on the high $p_0$ side) or $p_0/c\approx -2.6$ (on the low $p_0$ side)}
\label{s5}
	\end{center}
\end{figure}

This analysis provides a quantitative view of the correlation of depolarization with gap size, the existence of a gapped phase despite depolarization is visible in the raw data.  This is evident in Fig. \ref{s5}, where a finite incompressibility peak is visible after considerable polarization has been transferred.

\subsection{Single particle model}
 \begin{figure}[ht]
	\begin{center}
\includegraphics[width=.8\columnwidth]{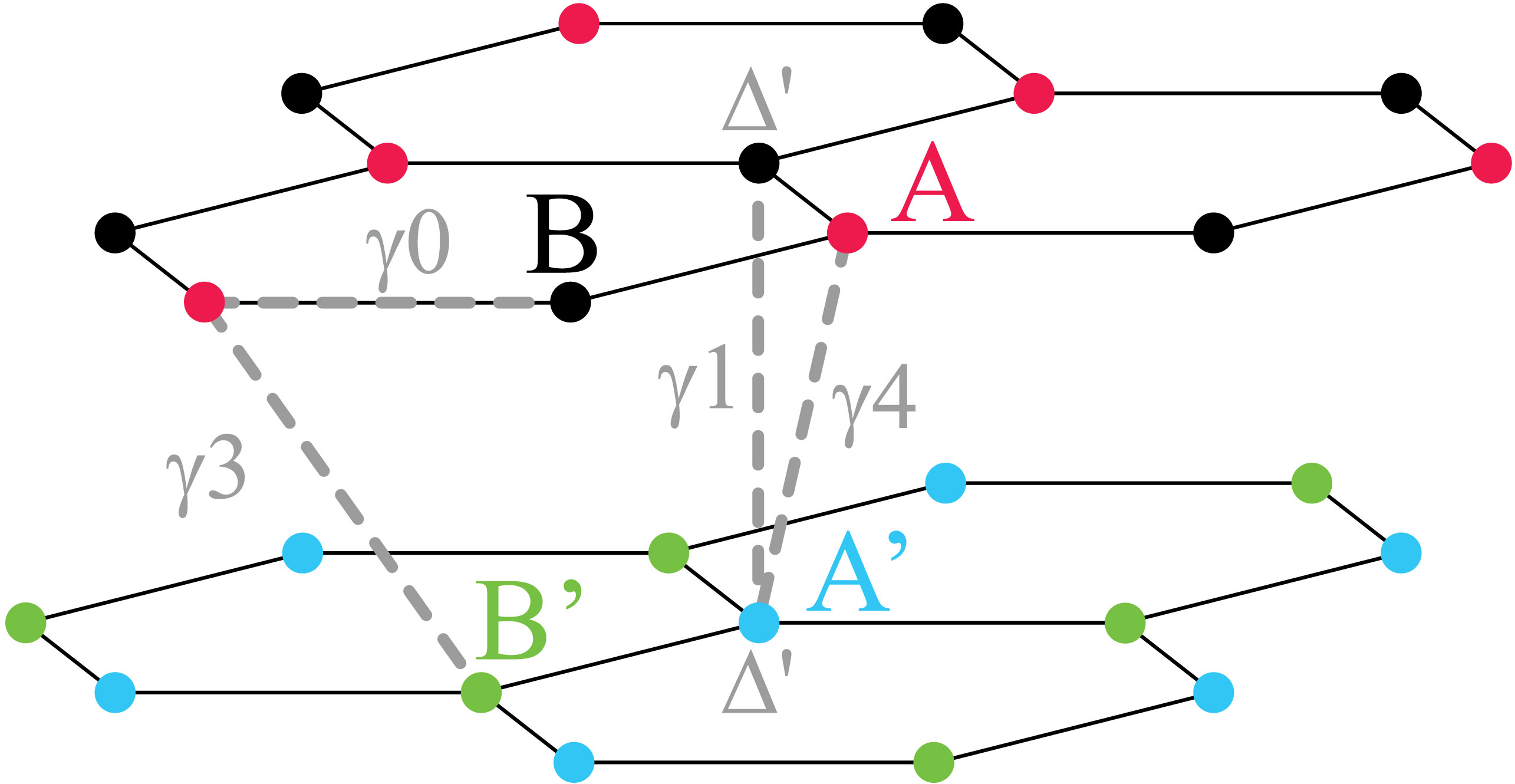}
		\caption{Bilayer graphene hopping parameters.}
		\label{fig:hopping}
	\end{center}
 \end{figure}
\begin{figure}[ht]
	\begin{center}
\includegraphics[width=2.5 in]{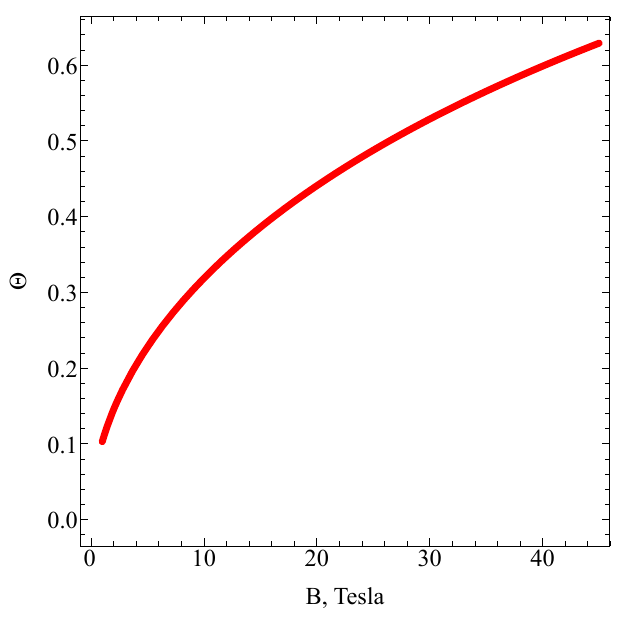}
		\caption{$\Theta$ as a function of $B$ within our tight binding model.}
		\label{fig:theta}
	\end{center}
 \end{figure}
\begin{figure}[ht]
\begin{center}
\includegraphics[width=.75 \columnwidth]{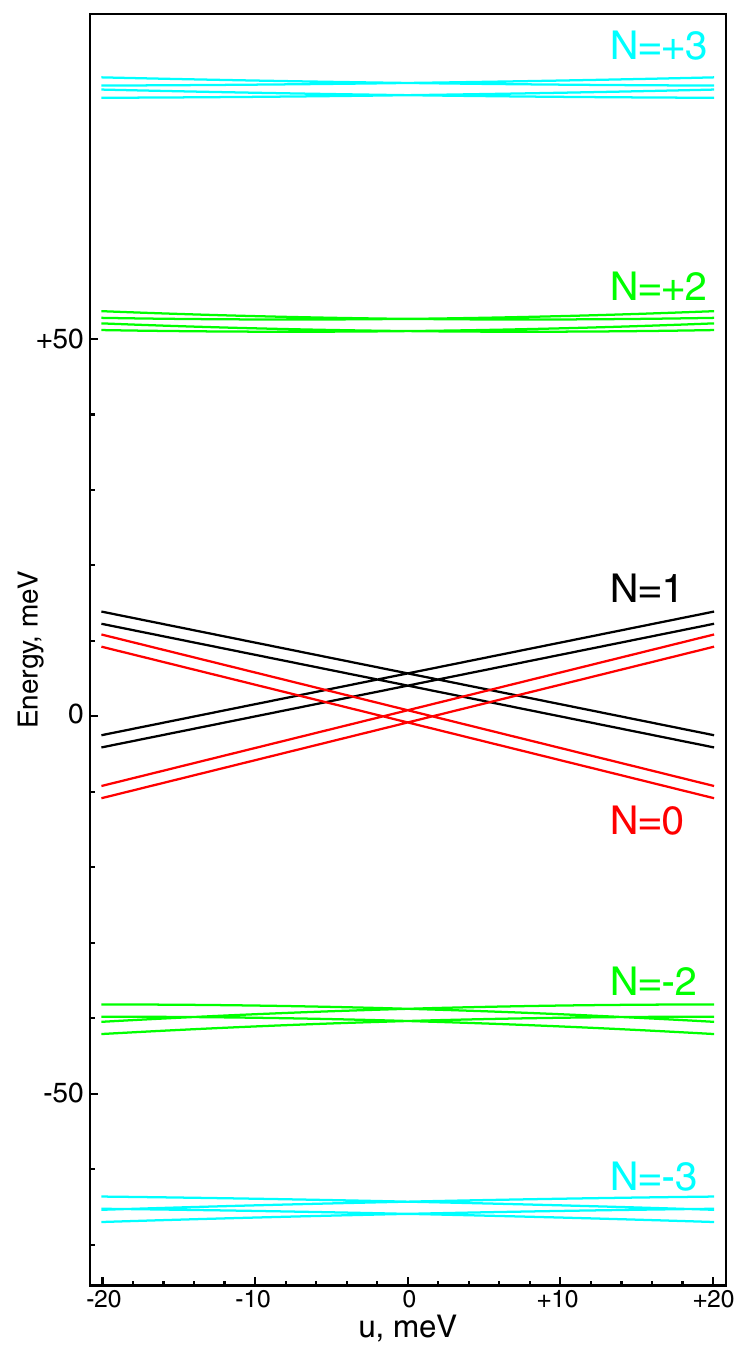} %183
\caption{Single particle energy spectrum of bilayer graphene at $B=14$T from a tight binding model. The Zeeman splitting is also included.  The interlayer potential difference $u\approx \frac{p_0}{c} \frac{c}{2c_0}\approx3.3\times 10^{-3} \frac{p_0}{c}$, does not exceed 15 meV in the data presented. For reference, the bare Coulomb energy at 14T is $e^2/\epsilon_{hBN}^\parallel \ell_B\approx 31$~meV, where we have taken $\epsilon_{hBN}^\parallel=6.6$\cite{geick_normal_1966}.  Interactions thus strongly mix the N=0 and N=1 levels (and in fact change the order of level filling within the ZLL, as shown in Ref. \onlinecite{hunt_competing_2016}), while mixing of the 0-1 manifold with the $|N|\geq2$ levels is comparatively weak.
}
\label{lls}
	\end{center}
\end{figure}

We use the same single-particle model for bilayer graphene as described in the supplementary information of \onlinecite{hunt_competing_2016}, which includes tight binding parameters $\gamma_0$, $\gamma_1$, $\gamma_4$, and $\Delta'$ (see Fig. \ref{fig:hopping}).
The single-particle  spectrum at $B=$14T is shown in Fig.~\ref{lls}, with a level ordering which is typical for all the $B$-fields used in the experiment.
Within this model, the ZLL wavefunctions in the different valleys (written in the lattice basis $\psi_{N\xi}=\left(\phi_A,\phi_{B'},\phi_{A'},\phi_B\right)$) are
  \begin{align}
\psi_{0+}&=
\left(
\begin{array}{c}
|0\rangle \\
0\\
0\\
0
\end{array}\right)&\psi_{1+}&=
\left(
\begin{array}{c}
\cos\Theta|1\rangle \\
0\\
\cos\Phi\sin\Theta|0\rangle\\
\sin\Phi\sin\Theta|0\rangle
\end{array}\right)
\\
\psi_{0-}&=
\left(
\begin{array}{c}
0\\
|0\rangle \\
0\\
0
\end{array}\right)&
\psi_{1-}&=
\left(
\begin{array}{c}
0\\
\cos\Theta|1\rangle \\
\sin\Phi\sin\Theta|0\rangle\\
\cos\Phi\sin\Theta|0\rangle
\end{array}\right)
\label{eq:ZLLwfs}
\end{align}
$|0\rangle$ and $|1\rangle$ denote the $n=0, 1$ magnetic oscillator states of a conventional parabolically dispersing system.
The layer polarization of the $|1\rangle$ orbitals is then $\alpha=\cos^2\Theta-\sin^2\Theta\left(\cos^2\Phi-\sin^2\Phi\right)$.
However, $\Phi$ is very small ($<.033$ for fields below 35T). Thus $\Phi$ does not shift the balance between $|0\rangle$ and $|1\rangle$ oscillator states, and so does not enter calculations of long range Coulomb effects such as fractional quantum Hall.  $\Theta$ controls the degree to which N=1 orbitals are strictly analogous to a parabolic electron system, i.e., purely composed of $|1\rangle$ oscillator states.  A plot of $\Theta(B)$ is shown in Fig. \ref{fig:theta}.

\clearpage
\subsection{Numerical simulations}

The phase diagram of Fig. 2d was calculated using the density matrix renormalization group (DMRG) for multicomponent quantum Hall systems\cite{white_density_1992, zaletel_infinite_2015}.
For a detailed exposition and justification of the approach used here, we refer to Ref.\onlinecite{hunt_competing_2016}, which benchmarked similar numerical computations against the experimentally determined layer polarization data.

The splitting $\Delta_{10}$ between the $N=0, 1$ orbitals of the ZLL is small, while the splitting between the ZLL and the higher LLs is large, of order $\hbar \omega_c$.
Our approach is based on the hierarchy $\Delta_{10} < E_C < \hbar \omega_c$, where $E_C$ is the Coulomb scale, which holds throughout the experimental range of $B$ and $p_0$ (see Fig \ref{lls}).
Because of the large  splitting between the ZLL and the $N \geq 2$ LLs, we project the problem into the ZLL, though will account for screening from the filled LLs through an effective interaction $V_{\textrm{eff}}$ we will discuss shortly. The ZLL projected Hamiltonian takes the form
\begin{widetext}
\begin{align}
H = \int d^2 q \, n_{\textrm{ZLL}}(q)V_{\textrm{eff}}(q) n_{\textrm{ZLL}}(-q) + \Delta_{10} \hat{N}_1 + \textrm{[isospin splittings]} + \textrm{[SU(4)-interaction anisotropies]}
\end{align}
\end{widetext}
Here $n_{\textrm{ZLL}}(q)$ is the Fourier transform of the 2D electron density projected into the ZLL, $V_{\textrm{eff}}$ is the effective interaction, $\hat{N}_1$ is the electron number in the $N=1$ orbital,  and $\Delta_{10}$ is the splitting between the $N=0, 1$ orbitals.
There are also  single-particle isospin (spin-valley) splittings and small SU(4) interaction anisotropies, but since our interest is in regions where the valence electrons partially occupy a single isospin, while all other isospins are either full or empty, these can be dropped from the problem.

	The bare Coulomb interaction is screened  by the encapsulating hBN, the graphite gates at a distance $d/2$ from the sample, and the filled LLs below the ZLL. Screening from the hBN is accounted for in the Coulomb scale $E_C = \frac{e^2}{4 \pi \ell_B \epsilon_{\textrm{BN}}}$, where $\epsilon_{\textrm{hBN}} \sim 6.6$ is the dielectric constant of the hBN\cite{geick_normal_1966}.
We assume the graphite behaves as a metal, so in units of  $\ell_B, E_C$, the gate screened interaction is $V(q) = \frac{2 \pi}{q} \tanh(q d)$.
The phase diagram here is presented for device A,  $d = 40$nm,  $d/\ell_B = 1.56 \sqrt{B/\textrm{T}}$, but the results are largely insensitive to $d$ at these fields since $d / \ell_B \gg 1$.

    The residual response of the filled LLs below the ZLL is controlled by the ratio $\frac{E_C}{\hbar \omega_c} \approx 0.5 - .8$.
We account for it within a phenomenological RPA-type treatment discussed in Refs.\cite{papic_topological_2014, hunt_competing_2016}, taking
\begin{align}
V_{\textrm{eff}}(q) = \frac{V(q)}{1 + a_{\textrm{scr}} V(q) \tanh(0.62 q^2 \ell_B^2) 4 \log(4)/ 2 \pi  }.
\end{align}
Within the two-band model of BLG at $\nu=0$, RPA calculations give $a_{\textrm{scr}} = \frac{E_C}{\hbar \omega_c} \equiv a_\ast$.
However, this value will not be quantitatively correct due to 4-band corrections and the filling of the isospins within the ZLL.
For this reason, we treat $a_{\textrm{scr}}$ as a phenomenological parameter. In Ref.\onlinecite{hunt_competing_2016}, quantitative agreement between numerics and experiment at 31T was obtained for $a_{\textrm{scr}} \sim 0.38 a_\ast$, which is the value used in the phase diagram of the main text.
To check that our conclusions are qualitatively insensitive  to screening,  the calculations below will be repeated for $a_{\textrm{scr}} = 0, 0.25 a_\ast, 0.5 a_\ast$ and $0.75 a_\ast$.

	We note that the RPA treatment only renormalizes the two-body Hamiltonian, while in principle higher-body terms are also generated.
However, the two-body screening diagrams are larger by a factor of $N_f = 4$ relative to three-body corrections, and three-body corrections actually vanish within the ``standard'' model of graphene, which only accounts for the nearest neighbor hoppings, due to particle-hole symmetry. When accounting for the further-neighbor hoppings, there is a small
amount of particle-hole symmetry breaking, but taken together this suggests the effective three-body interactions from outside the ZLL are much smaller than those which will be generated (and fully accounted for) from LL-mixing within the ZLL itself.

Under the assumption of isospin polarization, the $n_{\textrm{ZLL}}(q)$ can be restricted to the contribution from a single isospin, which contains the two LLs $N=0, 1$. In this study we keep the full Hilbert space of both, since the splitting $\Delta_{10}$ between them is small and mixing between them plays a crucial role in stabilizing the Pfaffian.
As discussed, within the four-band model of graphene, the $N=0$ orbital has the character of a conventional $n$=0 LL of GaAs, while the $N=1$ orbital  is an admixture of the conventional $n$=0 and $n$=1 LLs.
Letting $\bar{\rho}_{N N'}(q) \equiv \sum_{k} e^{-i k q_y \ell_B^2}c^\dagger_{N, k + q_x/2} c_{N', k - q_x/2}$ denote the ``guiding center'' density operator projected into orbitals $N, N'$,\cite{girvin_quantum_1987} the density operator is
\begin{align}
n_{\textrm{ZLL}}(q) = \sum_{N, N'} \mathcal{F}_{N, N'}(q)  \bar{\rho}_{N N'}(q)
\end{align}
where $\mathcal{F}$ are the BLG ``form factors'' for LL projection, which depend on the lattice structure of the wavefunctions, Eq.\ref{eq:ZLLwfs}.
Up to  small corrections at the lattice scale of order  $\frac{a_0}{\ell_B}$, the BLG form factors $\mathcal{F}_{MN}$ can be expressed as linear combinations of the conventional  form-factors $F_{mn}(q_x, q_y) =  e^{-q^2/4} \sqrt{ \frac{m!}{n!} } \left(\frac{q_x + i q_y}{\sqrt{2}} \right)^{n - m} L^{(n - m)}_{m}(q^2/2)$,
\begin{widetext}
\begin{align}
\mathcal{F}_{00} = F_{00}, \quad \mathcal{F}_{11} = \cos^2(\Theta) F_{11} + \sin^2( \Theta) F_{00}, \quad \mathcal{F}_{01} = \cos(\Theta) F_{01}.
\end{align}
\end{widetext}
where $\Theta$ is determined from the $B$-dependent band structure, Eq.\ref{eq:ZLLwfs}.
The two-band model is recovered in the limit $\Theta\rightarrow0$, while in the four-band model of bilayer graphene, $\Theta$ grows with perpendicular magnetic field $B$, Fig.\ref{fig:theta} \cite{hunt_competing_2016,jung_accurate_2014}.

\begin{figure*}[ht]
\begin{center}
\includegraphics[width=170mm]{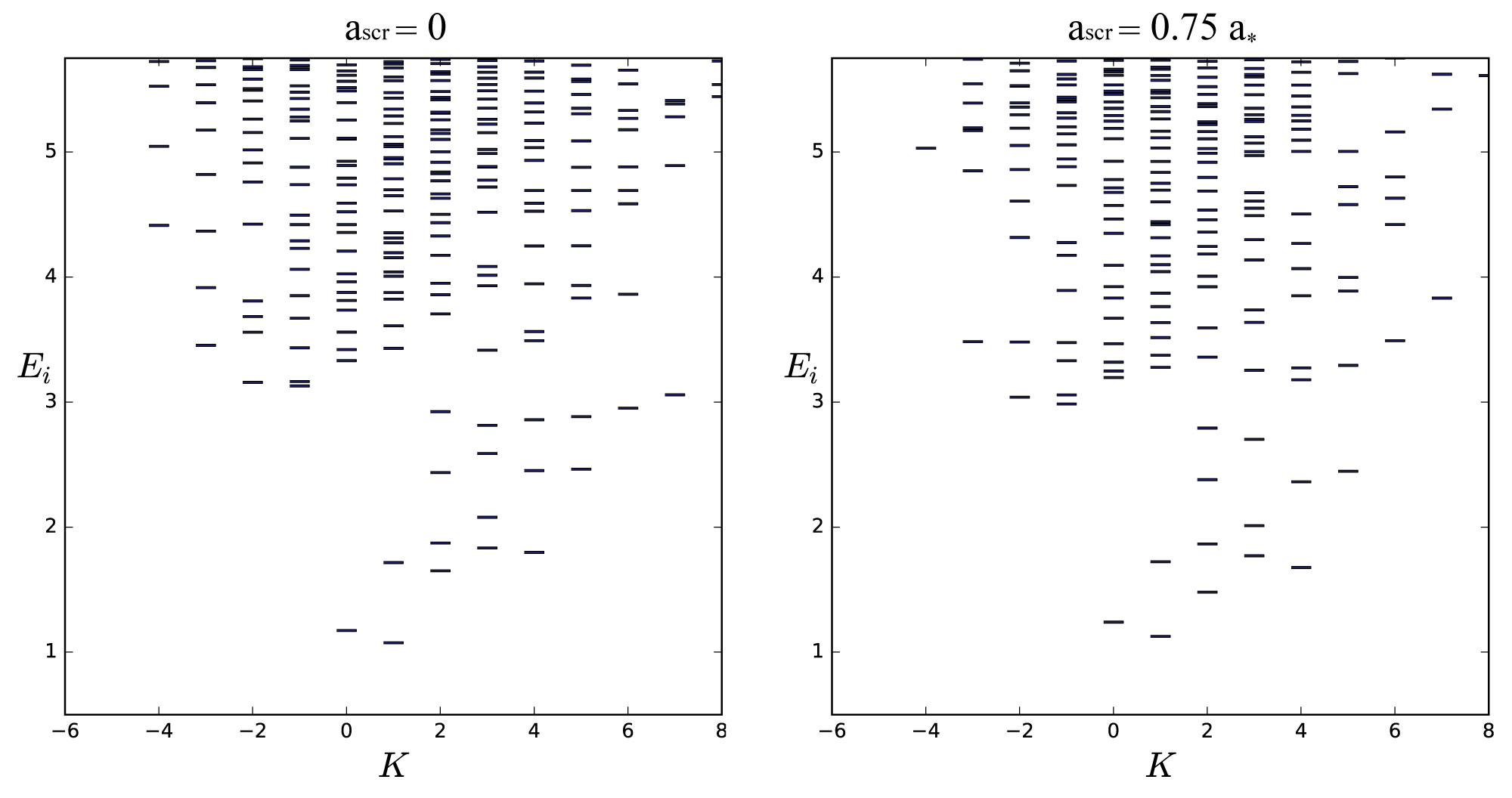} %183
\caption{Entanglement spectrum from cylinder DMRG simulations of the incompressible half-filled Landau level. $E_i$ is the spectrum of the reduced density matrix for one half of the cylinder, $E_i = -\log(\rho_L)$, plotted against the momentum $k$ around the cylinder.
The counting of low lying levels, $1, 2, 4, \cdots$, mimics the energy spectrum of the chiral edge theory of the Pfaffian phase, which we use to identify the state as the Pfaffian. The spectrum is shown for BLG model parameters $\Delta_{10} = 0.1 E_C, B = 12$T on a circumference $L = 19 \ell_B$ cylinder.
We repeat the calculations for two different values of the screening strength $a_{\textrm{scr}}$, where $a_\ast = \frac{E_C}{\hbar \omega_c}$, and find the Pfaffian is stable over a wide range.
}
\label{s4}
	\end{center}
\end{figure*}

Having defined the Hamiltonian, which depends on $B, \Delta_{10}/E_C$, and $a$, infinite-DMRG was used to obtain the ground state of an infinitely long, circumference $L_y = 19 \ell_B$ cylinder.
In the main text, the ground state was found on a 6$\times$16 grid of points in the $\Delta_{10}/E_C, B$ plane at screening strength $a_{\textrm{scr}} = 0.38 a_\ast$, keeping $m=12000$ DMRG states, with truncation errors less than  $10^{-6}$. The computations required around 32000 cpu-hours.
The Pfaffian ground state is identified by a finite correlation  length ($\xi \sim 3 \ell_B$ in the most robust regions) and distinctive entanglement spectrum characteristic of the associated edge chiral CFT \cite{li_Entanglement_2008,kitaev_topological_2006} (see Fig.~\ref{s4}).
The anti-Pfaffian, in contrast, would have an entanglement spectrum with the opposite chirality.
Unlike GaAs, scattering between the $N=0$ and $N=1$ orbitals strongly breaks particle-hole symmetry, unambiguously stabilizing the Pfaffian order over the particle-hole conjugated anti-Pfaffian.

Larger $\Theta$ (possibly achievable at the very highest $B$ fields) increases the correlation length of the Pfaffian and drives a transition into the compressible CFL phase, which is to be expected since sharper interactions favor the CFL phase.\cite{morf_transition_1998,rezayi_incompressible_2000,papic_tunable_2011}
Since the competing CFL phase is gapless, computing the precise location of the Pfaffian-CFL phase boundary is extremely difficult; finite size or finite entanglement effects turn the continuous transition into a crossover.
In particular, because  DMRG simulations at bond dimension $\chi$ can only capture at most $S = \log(\chi)$ entanglement, while the CFL has a log-divergent entanglement, it is impossible to exactly capture the CFL in DMRG. As discussed in detail in Ref.\onlinecite{geraedts_half-filled_2016}, the DMRG thus induces a ``finite-entanglement'' correlation length $\xi$. In the CFL phase, $\xi \sim \chi^\kappa$ diverges with increasing bond dimension, while in the Pfaffian phase, $\xi$ should converge to its physical value.
Furthermore, in the CFL, the entanglement should scale with the finite-entanglement correlation length as $S(\chi) = \frac{c}{6} \log(\xi(\chi)) + s_0$, were $c = 5$ at circumference $L = 19$.

\begin{figure*}[ht]
\begin{center}
\includegraphics[width=170mm]{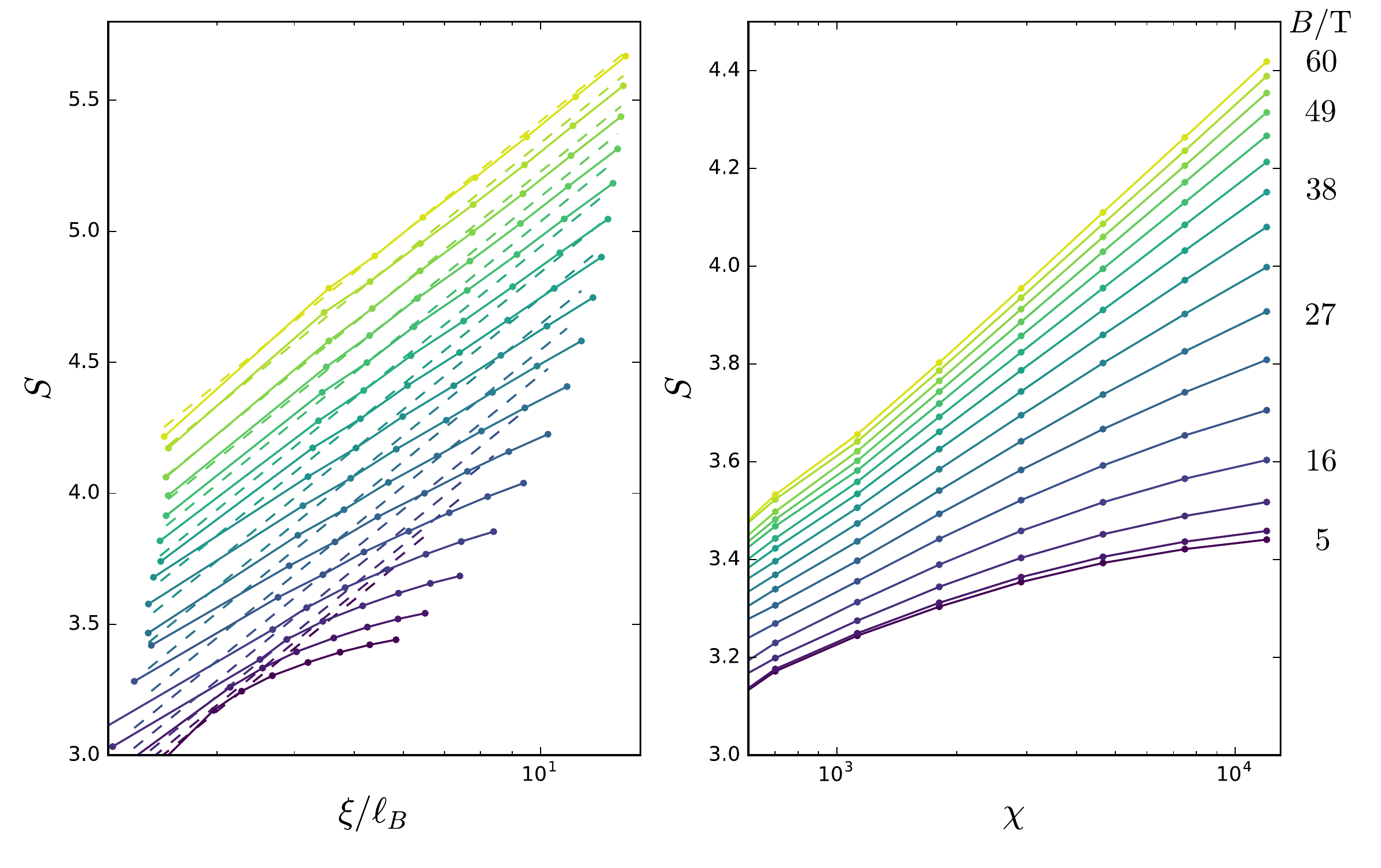} %183
\caption{Finite entanglement scaling for the Pfaffian to CFL transition. Data is shown for $L=19, a_{\textrm{scr}} = 0.5 a_\ast, \Delta_{10} / E_C = 0.5$ over a range of $B$. For each $B$, the DMRG simulations are repeated with increasing bond dimension (e.g., accuracy) $\chi = 600, \cdots , 12000$, measuring the bipartite entanglement $S$ and correlation length $\xi$  along the way. \textbf{Left}) $S(\xi)$, shown solid, is compared with the theoretical prediction for a CFL, $S = \frac{c}{6} \log(\xi) + s_0 , \,\, c = 5$, shown dashed. The log-linear scaling at high $B$ is consistent with a CFL, while the behavior at low $B$ is not, indicating a Pfaffian. \textbf{Right}) $S(\chi)$. For low $B$, the entanglement is converging with bond dimension, indicating the state is the gapped Pfaffian. For high $B$, the scaling is log-linear up to the largest $\chi$, consistent with the log-divergent entanglement of the CFL. Arguably the entanglement shows signs of saturating up $B \sim 34 - 38$T, beyond which we can't tell, suggesting the Pfaffian is stable up to at least this range.
}
\label{fig:FES}
	\end{center}
\end{figure*}

In Fig.~\ref{fig:FES}, we plot both $S(\chi)$ and $S(\xi)$ at $L = 19$, $\Delta_{10} / E_C = 0.5$, $a_{\textrm{scr}} = 0.5 a_\ast$ as the field increases from $B = 5 \cdots 60$T.
There is a regime where the simulations are clearly converging with $\chi$, indicating that the state is the gapped Pfaffian, and regions where the scaling is consistent with the CFL up to the largest $\chi$.
Precisely pinpointing the transition is clearly difficult, since we can't really distinguish between a Pfaffian phase of correlation length $\xi > 13$ and the CFL.

\begin{figure*}[ht]
\begin{center}
\includegraphics[width=170mm]{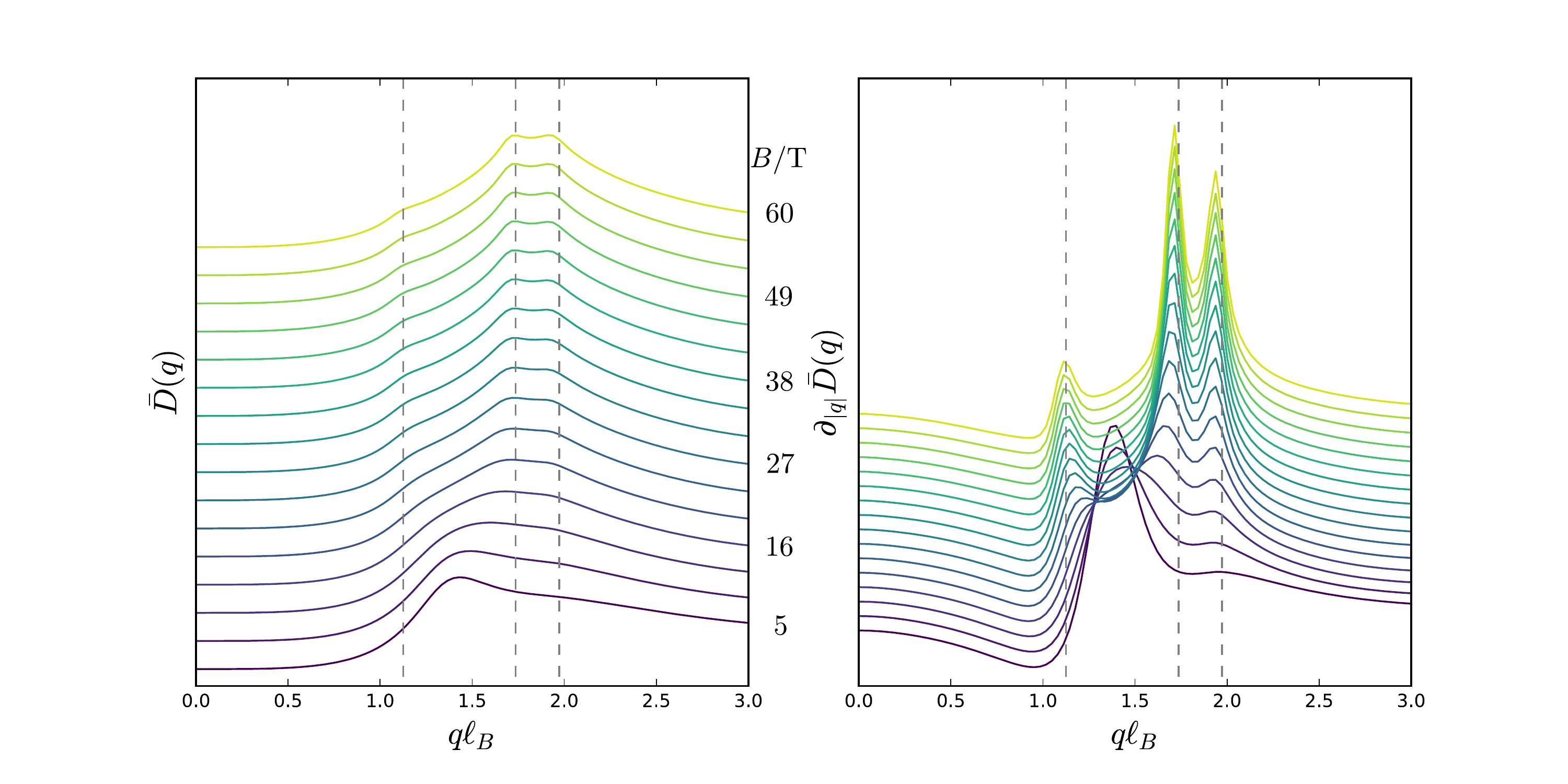} %183
\caption{The  guiding center density-density structure factor $\bar{D}(q) = \langle : n(q_x = q, q_y = 0) n(q_x = -q, q_y = 0) : \rangle$. The gray vertical lines indicate the location of predicted kinks for a perfectly circular composite Fermi surface.
At low $B$, where the state is a Pfaffian, there is only a single broad peak  unattached to these locations. At high $B$, where the state is transiting to a CFL, three sharp peaks develop. They are slightly displaced from the naive prediction because  the circular Fermi deforms slightly when it is placed on a cylinder.\cite{geraedts_half-filled_2016}
}
\label{fig:Dq}
\end{center}
\end{figure*}

Further insight is provided by the guiding-center density-density structure factor $\bar{D}(q)$ shown in Fig.~\ref{fig:Dq}. As discussed in Ref.\onlinecite{geraedts_half-filled_2016}, in the CFL phase $\bar{D}(q)$ is predicted to have non-analyticities at wave vectors associated with scattering a composite Fermion across the Fermi surface. At low $B$, there is only a broad bump, consistent with a Pfaffian, while at high $B$ three kinks develop close to the positions predicted for a circular Fermi surface. These kinks are brought into sharper relief by plotting ``$\partial_{|q|} \bar{D}(q)$,''\cite{geraedts_half-filled_2016} which is defined by multiplying the real-space correlations by $|x|$ before Fourier transforming to $q_x$.
Like the scaling of the entropy, the structure factor is consistent with a Pfaffian to CFL transition.

We note that a third competing phase is a striped phase, which breaks translation invariance. We have checked for this possibility by initializing the DMRG simulations with a charge density wave state close to the known wavelength of the stripe.\cite{rezayi_incompressible_2000} Throughout the phase diagram studied here, the stripes melt and form a liquid at sufficiently high DMRG accuracy. However, we do find that there is a critical $\Delta_{10} < 0$ where the Pfaffian phase is destroyed in favor of a stripe. Potentially, experiments could reach this regime by tuning $\Delta_{10}$ with very large electric fields (since the $N=0, 1$ orbitals have slightly different layer polarization), but none of our experiments are in this regime.

To address  the uncertainty in the screening strength,  we present calculations ($m$=6000, truncation error less than $10^{-5}$) for a variety of screening strengths $a_{\textrm{scr}}$.
Again, the Pfaffian is preferred throughout most of the phase diagram.
The screening has some quantitative effect on the growth of the correlation length, and hence presumably the precise location of the Pfaffian to CFL transition.

\begin{figure*}[ht]
\begin{center}
\includegraphics[width=170mm]{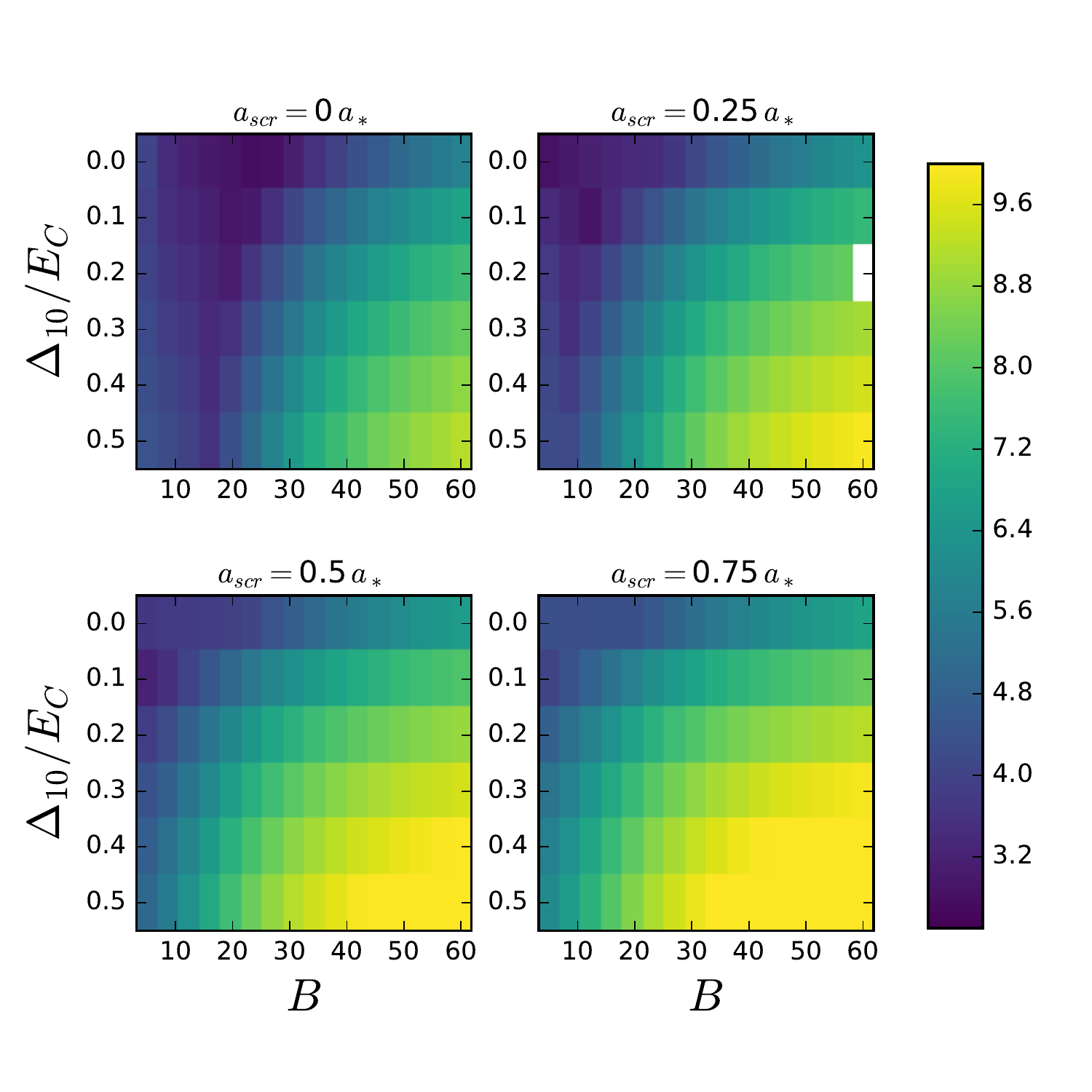} %183
\caption{The computed DMRG correlation length $\xi / \ell_B$ in the plane of $B, \Delta_{10}/E_C$ for several values of the screening strength $a$. The plots are analogous to that in the main text. Computations are on an $L = 19 \ell_B$ cylinder, with $\chi = 6000$ DMRG states. Larger screening increases the correlation length somewhat, but the general features are unchanged (e.g. the entanglement spectrum is consistent with the Pfaffian in the low-$\xi$ regime, Fig.\ref{s4})}
\label{fig:correlationlength}
\end{center}
\end{figure*}

While we computed  a phase diagram in the $\Delta_{10}/E_C, B$ plane, we do have a microscopic estimate of the splitting $\Delta_{10}$.
As discussed in detail in Ref.\cite{hunt_competing_2016}, it has three contributions: a $B$-dependent single particle splitting due to four-band corrections, a single particle splitting proportional the applied electric field $p_0$, and a ``Lamb-shift'' contribution arising from the exchange interaction with the levels below the ZLL.  The data from Sample C  shown in Fig. 2e were taken at $p_0$=6.0V. The trajectory shown in Fig. 2d is calculated including the effects of $B-$ and $p_0$ induced orbital splitting for $p_0=6V$ within the tight binding model described above, and also include the Lamb shift.
The $p_0$ dependence of $\Delta_{10}$ is an interesting feature, since it can be used to tune the strength of the LL-mixing in-situ.

\clearpage
\onecolumngrid
\appendix
\section{Supplementary Data Figures and Tables}
%
%\begin{itemize}
%  \item{Figure \ref{fig:comparison}:} Comparison between a metal gated and a graphite device.
%  \item{Figure \ref{moire}:} Landau fan for Device A, showing features associated with moire.
%  \item{Figure \ref{sz13}:} Capacitance and dissipation for Device A.
%  \item{Figure \ref{EMS13}:} Capacitance and dissipation for Device B.
%  \item{Figure \ref{HZS63}:} Capacitance and dissipation for Device C.
%  \item{Figure \ref{s9}:} Raw capacitance data in an N=1 level up to high magnetic field Sample C
%  \item{Figure \ref{s2}:} Plotted gaps in N=0 and N=1 LLs at 14T.
%  \item{Table \ref{table_gap_n0}:} Measured gaps in the N=0 LL for Sample A.
%    \item{Table \ref{table_gap_n1}:} Measured gaps in the N=1 LL for Sample A.
%\end{itemize}

\setcounter{figure}{0}
\setcounter{table}{0}
\renewcommand{\thefigure}{S\arabic{figure}}
\renewcommand{\thetable}{S\arabic{table}}
\renewcommand{\theequation}{S\arabic{equation}}

\begin{figure*}[ht]
\begin{center}
\includegraphics[width=183mm]{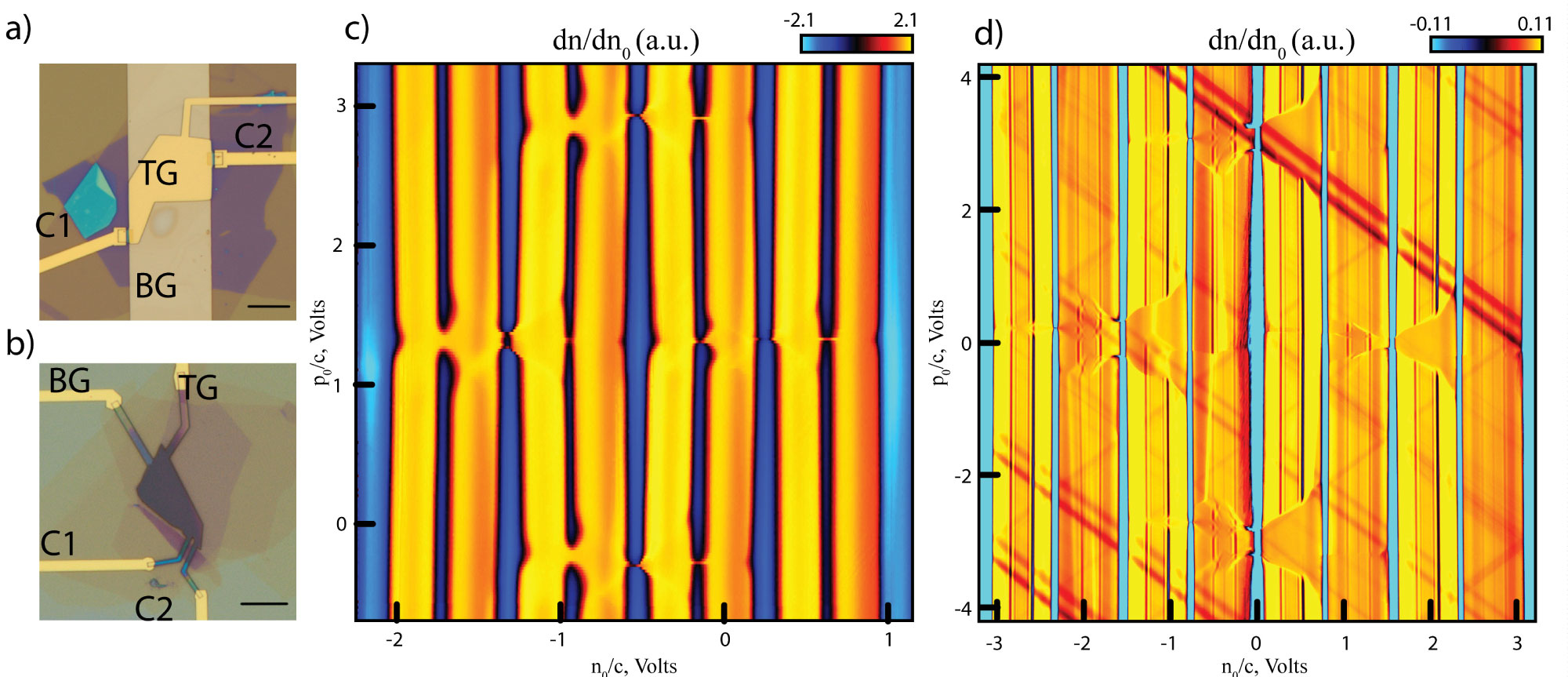} %%
%183
\caption{Comparison between metal gated (a) and graphite gated (b) devices (scale bar show 10um).  Symmetric capacitance, $C_S=C_T+C_B=\partial n/\partial n_0$, for a metal gated device at B=10 T (c).  The same measurement for a graphite gated device at B=12 T (d) shows much narrower integer QH states, along with many fractional states are visible.  Diagonal features in (d) are features that depend on only one of the gates (top or bottom) potentials $V_B$ or $V_T$, indicating either single gated regions or electronic structure within the graphite gates. }
\label{fig:comparison}
	\end{center}
\end{figure*}

\begin{figure*}[ht]
	\begin{center}
\includegraphics[width=89mm]{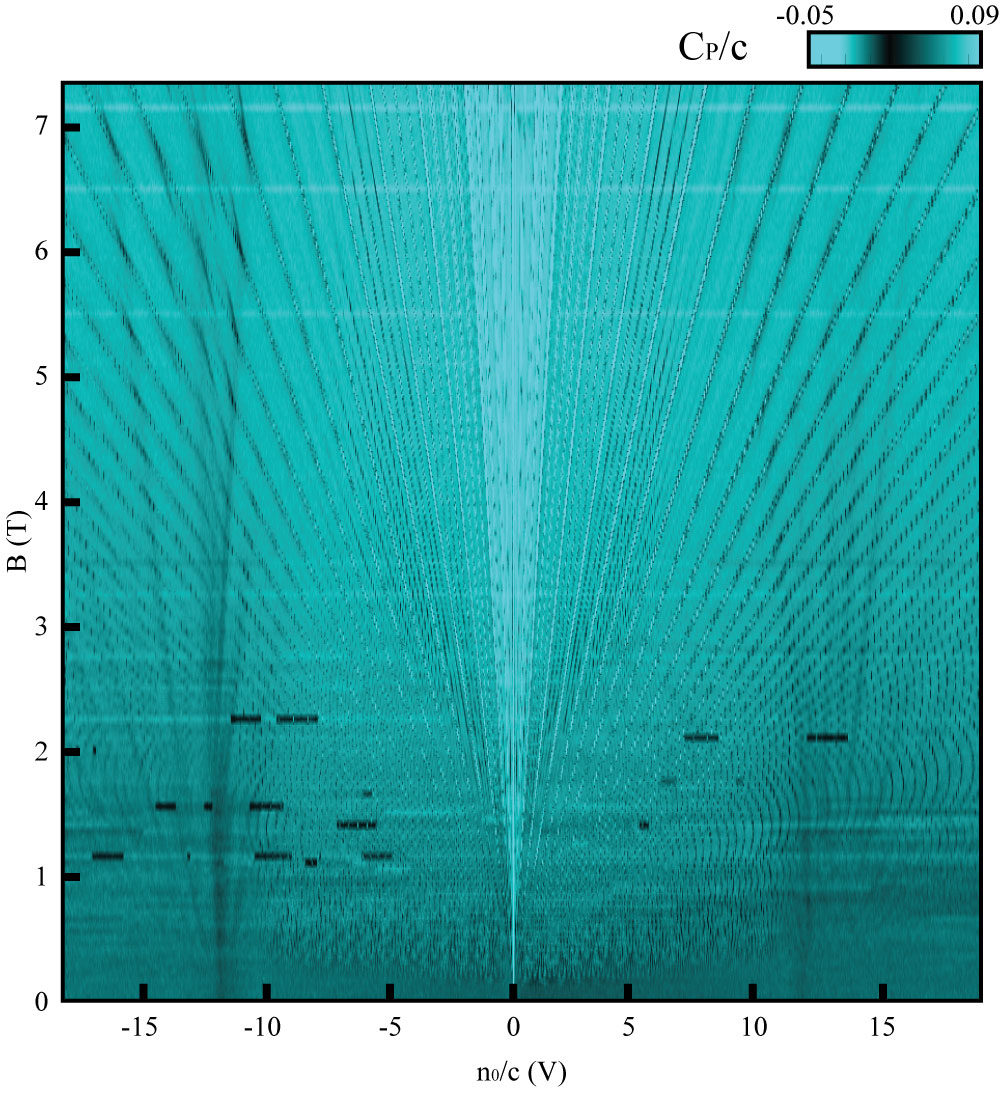} %183
\caption{Landau fan plot of $C_p/c$ in Sample A.  The main fan centered at charge neutrality is visible, along with satellite fans originating from superlattice zone boundary at $n0/c=\pm 11.8$~V.  The location of the satellite peaks imply a superlattice constant $\lambda \approx 13.3$ nm, corresponding to near-perfect (within experimental error) angular alignment of the Bernal bilayer to the underlying boron nitride.}
\label{moire}
	\end{center}
\end{figure*}

\begin{figure*}[ht]
\begin{center}
\includegraphics[width=183mm]{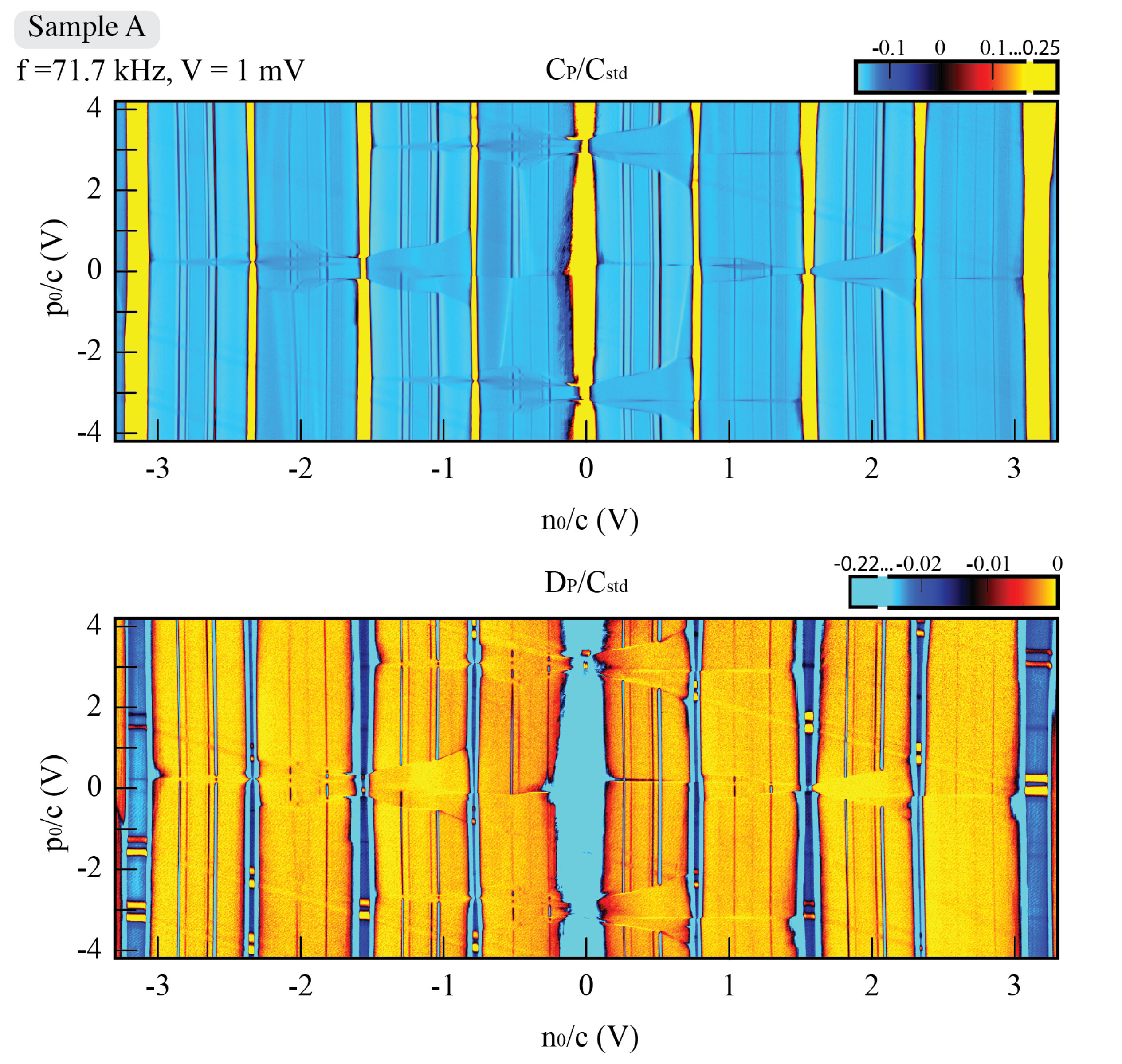} %183
\caption{Penetration field capacitance and associated dissipative signal for the data set shown in Figure 1 of the main text from Device A (\#SZ13), taken at base temperature and B=12T.  Data was taken with a 1 mV AC excitation at 71.77 kHz applied to the top gate.}
\label{sz13}
	\end{center}
\end{figure*}

\begin{figure*}[ht]
\begin{center}
\includegraphics[width=183mm]{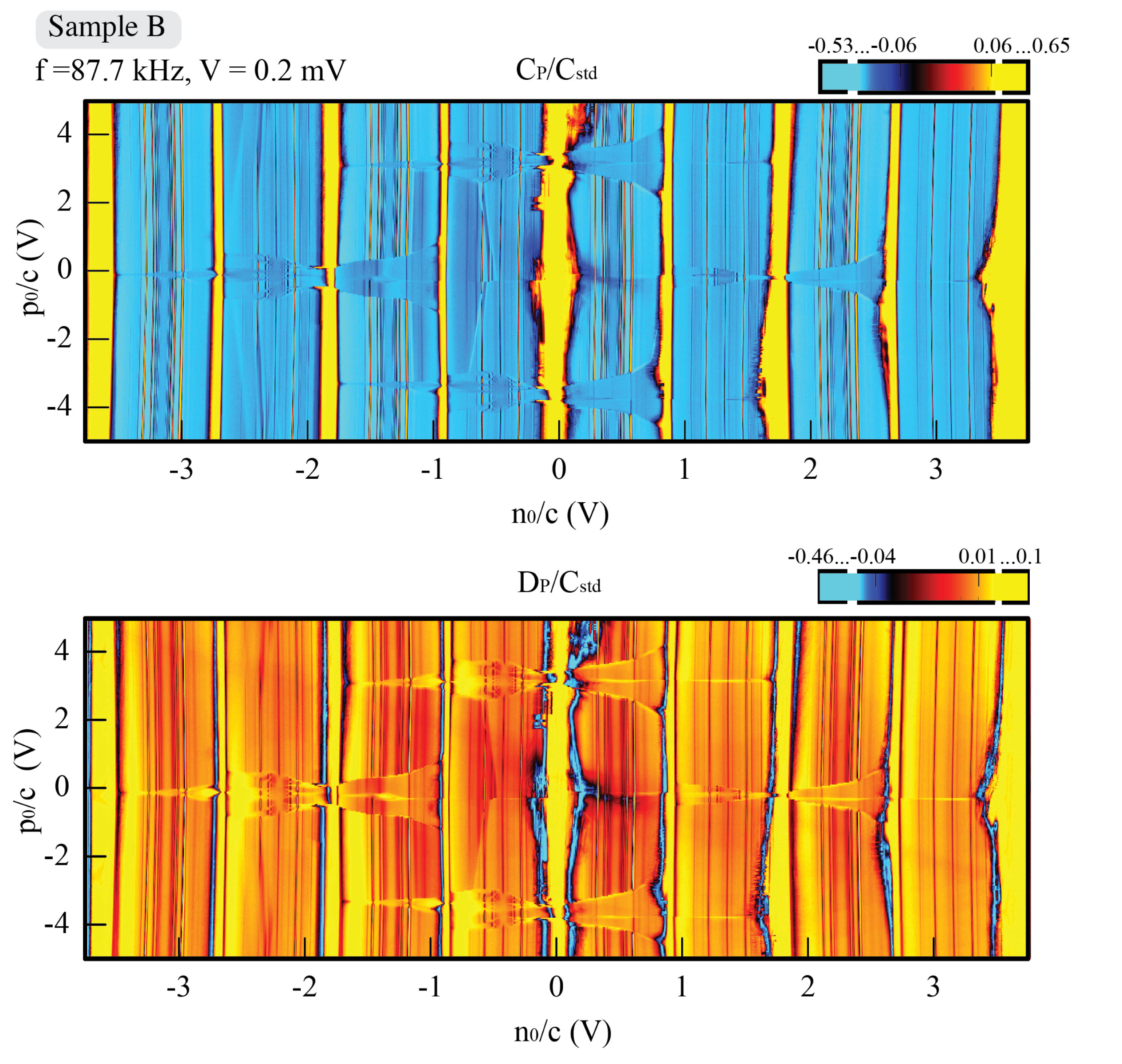} %183
\caption{Penetration field capacitance and associated dissipative signal for Device C (\#EMS13), taken at base temperature and B=14T.  Data was taken with a 0.2 mV AC excitation at 87.77 kHz applied to the top gate.}
\label{EMS13}
	\end{center}
\end{figure*}

\begin{figure*}[ht]
\begin{center}
\includegraphics[width=183mm]{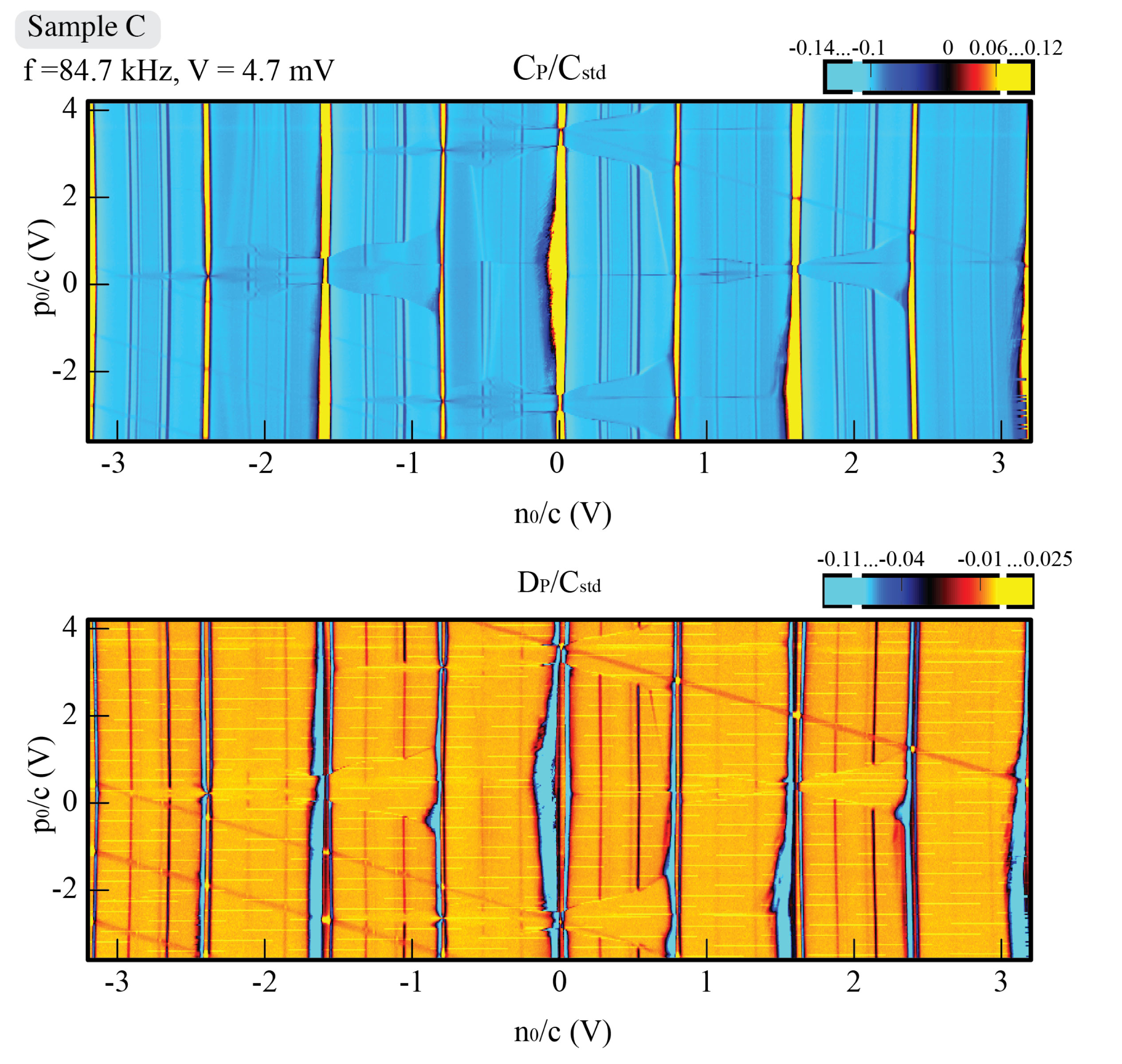} %183
\caption{Penetration field capacitance and associated dissipative signal for Device B (\#HZS63), taken at base temperature and B=14T.  Data was taken with a 4.7 mV AC excitation at 84.77 kHz applied to the top gate.}
\label{HZS63}
	\end{center}
\end{figure*}

\begin{figure*}[ht]
	\begin{center}
\includegraphics[width=100mm]{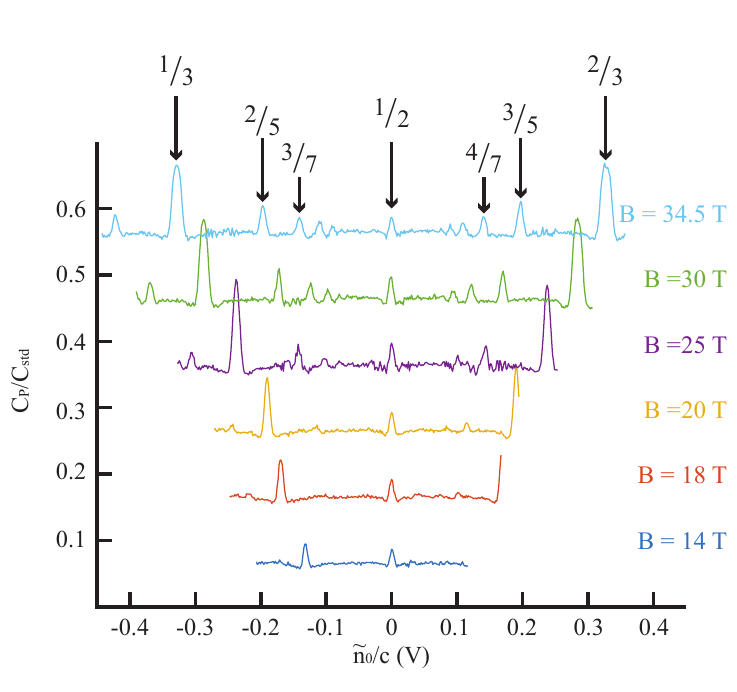} %183
\caption{Tuning of psuedopotentials in Sample C. Line-traces of penetration field capacitance $C_\text{pen}$ at different magnetic fields across $\nu=1\longrightarrow\nu=2$ Landau level at $p_0/c=6\text{ V}$. The line-traces are offset for visibility. As the magnetic field is increased the structure of the wavefunction is tuned from N=0-like to N=1-like.}
\label{s9}
	\end{center}
\end{figure*}

\clearpage
\begin{figure*}[ht]
	\begin{center}
\includegraphics[width=89mm]{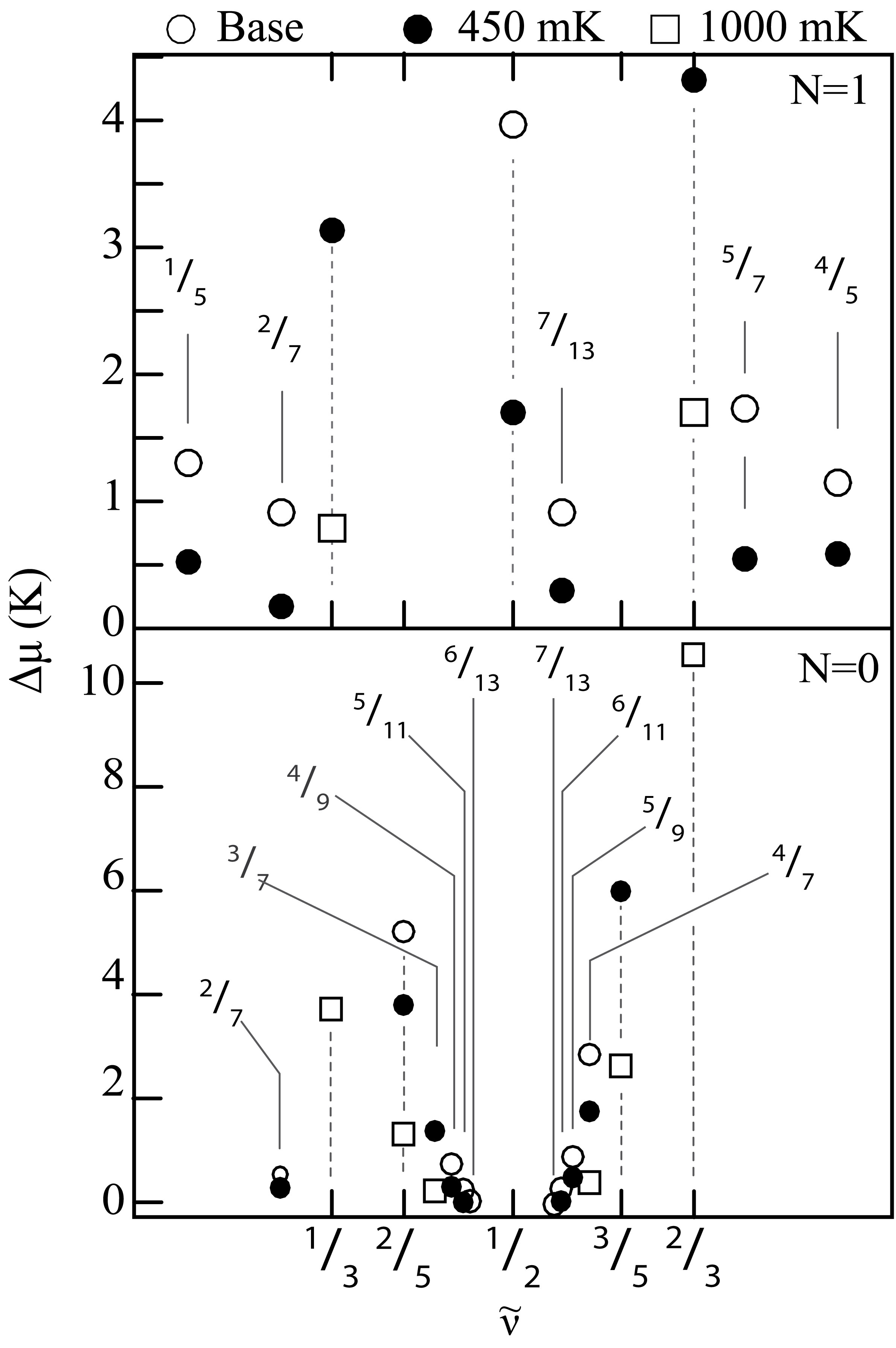} %183
\caption{Measured gaps in the N=0 and N=1 Landau levels in Sample A.  Gap measurements are taken at B=14 T, within one of the single component regimes between $\nu=0$ and $\nu=1$ (for the N=0 orbital) and $\nu=1$ and $\nu=2$ (for the N=1 orbital). Measurements are taken at temperatures from 20 mK to 1.1 K }
\label{s2}
	\end{center}
\end{figure*}

\begin{table}[ht]
\caption{Thermodynamic energy gaps in the single component N=0 orbital regime (Sample A).  Gaps are measured at B=14T and expressed in Kelvin.}
\label{table_gap_n0}
\begin{tabular}{|c|c|c|c|c|c|c|c|c|c|}
\hline                                          & \multicolumn{9}{|c|}{Temperature, mK}                            \\ \hline\cline{2-10}
$\tilde{\nu}$ & 20    & 160  & 310  & 450  & 577  & 711   & 810  & 1000 & 1150 \\ \hline
\multicolumn{1}{|c|}{$2/7$}                & 0.63  & 0.77 & 0.59 & 0.38 & 0.3  & 0.2   & -    & -    & -    \\ \hline \cline{1-1}
\multicolumn{1}{|c|}{$1/3$}                & -     & -    & -    & -    & -    & -     & -    & 3.77 & 3.1  \\ \hline \cline{1-1}
\multicolumn{1}{|c|}{2/5}                  & -     & -    & -    & 3.85 & 3.25 & 2.63  & 2.25 & 1.4  & 1.15 \\ \hline \cline{1-1}
\multicolumn{1}{|c|}{3/7}                  & -     & -    & 1.89 & 1.46 & 1.03 & 0.625 & 0.53 & 0.31 & -    \\ \hline\cline{1-1}
\multicolumn{1}{|c|}{4/9}                  & 0.83  & 0.8  & 0.62 & 0.40 & 0.29 & 0.15  & -    & -    & -    \\ \hline\cline{1-1}
\multicolumn{1}{|c|}{5/11}                 & 0.345 & 0.22 & 0.13 & 0.11 & -    & -     & -    & -    & -    \\ \hline\cline{1-1}
\multicolumn{1}{|c|}{6/13}                 & 0.12  & 0.07 & -    & -    & -    & -     & -    & -    & -    \\ \hline\cline{1-1}
\multicolumn{1}{|c|}{7/13}                 & 0.07  & 0.06 & 0.04 & -    & -    & -     & -    & -    & -    \\ \hline\cline{1-1}
\multicolumn{1}{|c|}{6/11}                 & 0.35  & 0.22 & 0.18 & 0.13 & -    & -     & -    & -    & -    \\ \hline\cline{1-1}
\multicolumn{1}{|c|}{5/9}                  & 0.96  & 0.98 & 0.79 & 0.57 & 0.38 & 0.2   & 0.18 & -    & -    \\ \hline\cline{1-1}
\multicolumn{1}{|c|}{4/7}                  & 2.9   & 2.67 & 2.34 & 1.83 & 1.36 & 0.87  & 0.89 & 0.48 & 0.34 \\ \hline \cline{1-1}
\multicolumn{1}{|c|}{3/5}                  & -     & -    & -    & 6.0  & 5.23 & 4.36  & 3.8  & 2.7  & 2.1  \\ \hline \cline{1-1}
\multicolumn{1}{|c|}{2/3}                  & -     & -    & -    & -    & -    & -     & -    & 10.5 & 9.56 \\ \hline
\end{tabular}
\end{table}

\begin{table}[ht]
\caption{Thermodynamic energy gaps in the single component N=1 orbital regime.  Gaps are measured at B=14T and expressed in Kelvin.}
\label{table_gap_n1}
\begin{tabular}{|c|c|c|c|c|c|c|c|c|c|}
\hline & \multicolumn{9}{|c|}{Temperature, mK}                          \\ \hline\cline{2-10}
\multicolumn{1}{|c|}{$\tilde{\nu}$} & 20   & 160  & 310  & 450  & 577  & 711  & 810  & 1000 & 1150 \\ \hline
\multicolumn{1}{|c|}{1/5}                  & 1.3  & 0.9  & 0.94 & 0.51 & 0.3  & -    & -    & -    & -    \\\hline \cline{1-1}
\multicolumn{1}{|c|}{2/7}                  & 0.9  & 0.39 & 0.33 & 0.15 & 0.1  & -    & -    & -    & -    \\\hline \cline{1-1}
\multicolumn{1}{|c|}{1/3}                  & -    & -    & -    & 3.15 & 2.94 & 2.4  & 1.76 & 0.78 & 0.63 \\\hline \cline{1-1}
\multicolumn{1}{|c|}{1/2}                  & 3.99 & 3.17 & 2.68 & 1.7  & 1.01 & 0.57 & 0.46 & -    & -    \\\hline \cline{1-1}
\multicolumn{1}{|c|}{7/13}                 & 0.9  & 0.44 & 0.4  & 0.28 & -    & -    & -    & -    & -    \\\hline \cline{1-1}
\multicolumn{1}{|c|}{2/3}                  & -    & -    & -    & 4.35 & 3.79 & 3.12 & 2.6  & 1.7  & 1.1  \\ \hline\cline{1-1}
\multicolumn{1}{|c|}{5/7}                  & 1.73 & 1.02 & 0.8  & 0.53 & 0.43 & 0.2  & 0.18 & -    & -    \\\hline \cline{1-1}
\multicolumn{1}{|c|}{4/5}                  & 1.14 & 0.98 & 0.81 & 0.57 & 0.43 & 0.21 & 0.16 & -    & -   \\ \hline
\end{tabular}
\end{table}


\clearpage

\begin{thebibliography}{10}
\expandafter\ifx\csname url\endcsname\relax
  \def\url#1{\texttt{#1}}\fi
\expandafter\ifx\csname urlprefix\endcsname\relax\def\urlprefix{URL }\fi
\providecommand{\bibinfo}[2]{#2}
\providecommand{\eprint}[2][]{\url{#2}}

\bibitem{kitaev_fault-tolerant_2003}
\bibinfo{author}{Kitaev, A.~Y.}
\newblock \bibinfo{title}{Fault-tolerant quantum computation by anyons}.
\newblock \emph{\bibinfo{journal}{Annals of Physics}}
  \textbf{\bibinfo{volume}{303}}, \bibinfo{pages}{2--30}
  (\bibinfo{year}{2003}).

\bibitem{nayak_non-abelian_2008}
\bibinfo{author}{Nayak, C.}, \bibinfo{author}{Simon, S.~H.},
  \bibinfo{author}{Stern, A.}, \bibinfo{author}{Freedman, M.} \&
  \bibinfo{author}{Das~Sarma, S.}
\newblock \bibinfo{title}{Non-{Abelian} anyons and topological quantum
  computation}.
\newblock \emph{\bibinfo{journal}{Rev. Mod. Phys.}}
  \textbf{\bibinfo{volume}{80}}, \bibinfo{pages}{1083--1159}
  (\bibinfo{year}{2008}).

\bibitem{willett_observation_1987}
\bibinfo{author}{Willett, R.} \emph{et~al.}
\newblock \bibinfo{title}{Observation of an even-denominator quantum number in
  the fractional quantum {Hall} effect}.
\newblock \emph{\bibinfo{journal}{Phys. Rev. Lett.}}
  \textbf{\bibinfo{volume}{59}} (\bibinfo{year}{1987}).

\bibitem{ki_observation_2014}
\bibinfo{author}{Ki, D.-K.}, \bibinfo{author}{Fal’ko, V.~I.},
  \bibinfo{author}{Abanin, D.~A.} \& \bibinfo{author}{Morpurgo, A.~F.}
\newblock \bibinfo{title}{Observation of {Even} {Denominator} {Fractional}
  {Quantum} {Hall} {Effect} in {Suspended} {Bilayer} {Graphene}}.
\newblock \emph{\bibinfo{journal}{Nano Letters}} \textbf{\bibinfo{volume}{14}},
  \bibinfo{pages}{2135--2139} (\bibinfo{year}{2014}).

\bibitem{falson_even-denominator_2015}
\bibinfo{author}{Falson, J.} \emph{et~al.}
\newblock \bibinfo{title}{Even-denominator fractional quantum {Hall} physics in
  {ZnO}}.
\newblock \emph{\bibinfo{journal}{Nature Physics}}
  \textbf{\bibinfo{volume}{11}}, \bibinfo{pages}{347--351}
  (\bibinfo{year}{2015}).

\bibitem{moore_nonabelions_1991}
\bibinfo{author}{Moore, G.} \& \bibinfo{author}{Read, N.}
\newblock \bibinfo{title}{Nonabelions in the fractional quantum {Hall} effect}.
\newblock \emph{\bibinfo{journal}{Nuclear Physics B}}
  \textbf{\bibinfo{volume}{360}}, \bibinfo{pages}{362--396}
  (\bibinfo{year}{1991}).

\bibitem{eisenstein_compressibility_1994}
\bibinfo{author}{Eisenstein, J.~P.}, \bibinfo{author}{Pfeiffer, L.~N.} \&
  \bibinfo{author}{West, K.~W.}
\newblock \bibinfo{title}{Compressibility of the two-dimensional electron gas:
  {Measurements} of the zero-field exchange energy and fractional quantum
  {Hall} gap}.
\newblock \emph{\bibinfo{journal}{Phys. Rev. B}} \textbf{\bibinfo{volume}{50}},
  \bibinfo{pages}{1760--1778} (\bibinfo{year}{1994}).

\bibitem{cooper_observable_2009}
\bibinfo{author}{Cooper, N.~R.} \& \bibinfo{author}{Stern, A.}
\newblock \bibinfo{title}{Observable {Bulk} {Signatures} of {Non}-{Abelian}
  {Quantum} {Hall} {States}}.
\newblock \emph{\bibinfo{journal}{Physical Review Letters}}
  \textbf{\bibinfo{volume}{102}}, \bibinfo{pages}{176807}
  (\bibinfo{year}{2009}).

\bibitem{papic_tunable_2011}
\bibinfo{author}{Papic, Z.}, \bibinfo{author}{Abanin, D.~A.},
  \bibinfo{author}{Barlas, Y.} \& \bibinfo{author}{Bhatt, R.~N.}
\newblock \bibinfo{title}{Tunable interactions and phase transitions in {Dirac}
  materials in a magnetic field}.
\newblock \emph{\bibinfo{journal}{Phys. Rev. B}} \textbf{\bibinfo{volume}{84}}
  (\bibinfo{year}{2011}).

\bibitem{metlitski_cooper_2015}
\bibinfo{author}{Metlitski, M.~A.}, \bibinfo{author}{Mross, D.~F.},
  \bibinfo{author}{Sachdev, S.} \& \bibinfo{author}{Senthil, T.}
\newblock \bibinfo{title}{Cooper pairing in non-{Fermi} liquids}.
\newblock \emph{\bibinfo{journal}{Physical Review B}}
  \textbf{\bibinfo{volume}{91}}, \bibinfo{pages}{115111}
  (\bibinfo{year}{2015}).

\bibitem{liu_evolution_2011}
\bibinfo{author}{Liu, Y.} \emph{et~al.}
\newblock \bibinfo{title}{Evolution of the 7/2 {Fractional} {Quantum} {Hall}
  {State} in {Two}-{Subband} {Systems}}.
\newblock \emph{\bibinfo{journal}{Phys. Rev. Lett.}}
  \textbf{\bibinfo{volume}{107}} (\bibinfo{year}{2011}).

\bibitem{jain_composite-fermion_1989}
\bibinfo{author}{Jain, J.~K.}
\newblock \bibinfo{title}{Composite-fermion approach for the fractional quantum
  {Hall} effect}.
\newblock \emph{\bibinfo{journal}{Phys. Rev. Lett.}}
  \textbf{\bibinfo{volume}{63}} (\bibinfo{year}{1989}).

\bibitem{halperin_theory_1993}
\bibinfo{author}{Halperin, B.~I.}, \bibinfo{author}{Lee, P.~A.} \&
  \bibinfo{author}{Read, N.}
\newblock \bibinfo{title}{Theory of the half-filled {Landau} level}.
\newblock \emph{\bibinfo{journal}{Phys. Rev. B}} \textbf{\bibinfo{volume}{47}},
  \bibinfo{pages}{7312--7343} (\bibinfo{year}{1993}).

\bibitem{willett_experimental_1993}
\bibinfo{author}{Willett, R.~L.}, \bibinfo{author}{Ruel, R.~R.},
  \bibinfo{author}{West, K.~W.} \& \bibinfo{author}{Pfeiffer, L.~N.}
\newblock \bibinfo{title}{Experimental demonstration of a {Fermi} surface at
  one-half filling of the lowest {Landau} level}.
\newblock \emph{\bibinfo{journal}{Physical Review Letters}}
  \textbf{\bibinfo{volume}{71}}, \bibinfo{pages}{3846--3849}
  (\bibinfo{year}{1993}).

\bibitem{kang_how_1993}
\bibinfo{author}{Kang, W.}, \bibinfo{author}{Stormer, H.~L.},
  \bibinfo{author}{Pfeiffer, L.~N.}, \bibinfo{author}{Baldwin, K.~W.} \&
  \bibinfo{author}{West, K.~W.}
\newblock \bibinfo{title}{How real are composite fermions?}
\newblock \emph{\bibinfo{journal}{Physical Review Letters}}
  \textbf{\bibinfo{volume}{71}}, \bibinfo{pages}{3850--3853}
  (\bibinfo{year}{1993}).

\bibitem{read_paired_2000}
\bibinfo{author}{Read, N.} \& \bibinfo{author}{Green, D.}
\newblock \bibinfo{title}{Paired states of fermions in two dimensions with
  breaking of parity and time-reversal symmetries and the fractional quantum
  {Hall} effect}.
\newblock \emph{\bibinfo{journal}{Physical Review B}}
  \textbf{\bibinfo{volume}{61}}, \bibinfo{pages}{10267--10297}
  (\bibinfo{year}{2000}).

\bibitem{halperin_theory_1983}
\bibinfo{author}{Halperin, B.~I.}
\newblock \bibinfo{title}{Theory of the quantized {Hall} conductance}.
\newblock \emph{\bibinfo{journal}{Helv. Phys. Acta}}
  \textbf{\bibinfo{volume}{56}}, \bibinfo{pages}{75--102}
  (\bibinfo{year}{1983}).

\bibitem{lee_chemical_2014}
\bibinfo{author}{Lee, K.} \emph{et~al.}
\newblock \bibinfo{title}{Chemical potential and quantum {Hall} ferromagnetism
  in bilayer graphene}.
\newblock \emph{\bibinfo{journal}{Science}} \textbf{\bibinfo{volume}{345}},
  \bibinfo{pages}{58--61} (\bibinfo{year}{2014}).

\bibitem{maher_tunable_2014}
\bibinfo{author}{Maher, P.} \emph{et~al.}
\newblock \bibinfo{title}{Tunable fractional quantum {Hall} phases in bilayer
  graphene}.
\newblock \emph{\bibinfo{journal}{Science}} \textbf{\bibinfo{volume}{345}},
  \bibinfo{pages}{61--64} (\bibinfo{year}{2014}).

\bibitem{hunt_competing_2016}
\bibinfo{author}{Hunt, B.~M.} \emph{et~al.}
\newblock \bibinfo{title}{Competing valley, spin, and orbital symmetry breaking
  in bilayer graphene}.
\newblock \emph{\bibinfo{journal}{arXiv:1607.06461}}  (\bibinfo{year}{2016}).

\bibitem{levin_particle-hole_2007}
\bibinfo{author}{Levin, M.}, \bibinfo{author}{Halperin, B.~I.} \&
  \bibinfo{author}{Rosenow, B.}
\newblock \bibinfo{title}{Particle-{Hole} {Symmetry} and the {Pfaffian}
  {State}}.
\newblock \emph{\bibinfo{journal}{Physical Review Letters}}
  \textbf{\bibinfo{volume}{99}}, \bibinfo{pages}{236806}
  (\bibinfo{year}{2007}).

\bibitem{lee_particle-hole_2007}
\bibinfo{author}{Lee, S.-S.}, \bibinfo{author}{Ryu, S.},
  \bibinfo{author}{Nayak, C.} \& \bibinfo{author}{Fisher, M. P.~A.}
\newblock \bibinfo{title}{Particle-{Hole} {Symmetry} and the ν=5/2 {Quantum}
  {Hall} {State}}.
\newblock \emph{\bibinfo{journal}{Physical Review Letters}}
  \textbf{\bibinfo{volume}{99}}, \bibinfo{pages}{236807}
  (\bibinfo{year}{2007}).

\bibitem{kumar_nonconventional_2010}
\bibinfo{author}{Kumar, A.}, \bibinfo{author}{Csathy, G.~A.},
  \bibinfo{author}{Manfra, M.~J.}, \bibinfo{author}{Pfeiffer, L.~N.} \&
  \bibinfo{author}{West, K.~W.}
\newblock \bibinfo{title}{Nonconventional {Odd}-{Denominator} {Fractional}
  {Quantum} {Hall} {States} in the {Second} {Landau} {Level}}.
\newblock \emph{\bibinfo{journal}{Phys. Rev. Lett.}}
  \textbf{\bibinfo{volume}{105}} (\bibinfo{year}{2010}).

\bibitem{apalkov_stable_2011}
\bibinfo{author}{Apalkov, V.~M.} \& \bibinfo{author}{Chakraborty, T.}
\newblock \bibinfo{title}{Stable {Pfaffian} {State} in {Bilayer} {Graphene}}.
\newblock \emph{\bibinfo{journal}{Phys. Rev. Lett.}}
  \textbf{\bibinfo{volume}{107}} (\bibinfo{year}{2011}).

\bibitem{papic_topological_2014}
\bibinfo{author}{Papic, Z.} \& \bibinfo{author}{Abanin, D.~A.}
\newblock \bibinfo{title}{Topological {Phases} in the {Zeroth} {Landau} {Level}
  of {Bilayer} {Graphene}}.
\newblock \emph{\bibinfo{journal}{Physical Review Letters}}
  \textbf{\bibinfo{volume}{112}}, \bibinfo{pages}{046602}
  (\bibinfo{year}{2014}).

\bibitem{rezayi_breaking_2011}
\bibinfo{author}{Rezayi, E.~H.} \& \bibinfo{author}{Simon, S.~H.}
\newblock \bibinfo{title}{Breaking of {Particle}-{Hole} {Symmetry} by {Landau}
  {Level} {Mixing} in the ν=5/2 {Quantized} {Hall} {State}}.
\newblock \emph{\bibinfo{journal}{Physical Review Letters}}
  \textbf{\bibinfo{volume}{106}}, \bibinfo{pages}{116801}
  (\bibinfo{year}{2011}).

\bibitem{zaletel_infinite_2015}
\bibinfo{author}{Zaletel, M.~P.}, \bibinfo{author}{Mong, R. S.~K.},
  \bibinfo{author}{Pollmann, F.} \& \bibinfo{author}{Rezayi, E.~H.}
\newblock \bibinfo{title}{Infinite density matrix renormalization group for
  multicomponent quantum {Hall} systems}.
\newblock \emph{\bibinfo{journal}{Physical Review B}}
  \textbf{\bibinfo{volume}{91}} (\bibinfo{year}{2015}).
\newblock \bibinfo{note}{ArXiv: 1410.3861}.

\bibitem{rezayi_landau-level-mixing_2017}
\bibinfo{author}{Rezayi, E.~H.}
\newblock \bibinfo{title}{Landau-level-mixing and the ground state of the 5/2
  quantum {Hall} effect}.
\newblock \emph{\bibinfo{journal}{arXiv:1704.03026 [cond-mat]}}
  (\bibinfo{year}{2017}).
\newblock \bibinfo{note}{ArXiv: 1704.03026}.

\bibitem{levin_collective_2009}
\bibinfo{author}{Levin, M.} \& \bibinfo{author}{Halperin, B.~I.}
\newblock \bibinfo{title}{Collective states of non-{Abelian} quasiparticles in
  a magnetic field}.
\newblock \emph{\bibinfo{journal}{Physical Review B}}
  \textbf{\bibinfo{volume}{79}}, \bibinfo{pages}{205301}
  (\bibinfo{year}{2009}).

\bibitem{von_keyserlingk_enhanced_2015}
\bibinfo{author}{von Keyserlingk, C.}, \bibinfo{author}{Simon, S.} \&
  \bibinfo{author}{Rosenow, B.}
\newblock \bibinfo{title}{Enhanced {Bulk}-{Edge} {Coulomb} {Coupling} in
  {Fractional} {Fabry}-{Perot} {Interferometers}}.
\newblock \emph{\bibinfo{journal}{Physical Review Letters}}
  \textbf{\bibinfo{volume}{115}}, \bibinfo{pages}{126807}
  (\bibinfo{year}{2015}).

\bibitem{wei_mach-zehnder_2017}
\bibinfo{author}{Wei, D.~S.} \emph{et~al.}
\newblock \bibinfo{title}{Mach-{Zehnder} interferometry using spin- and
  valley-polarized quantum {Hall} edge states in graphene}.
\newblock \emph{\bibinfo{journal}{arXiv:1703.00110 [cond-mat]}}
  (\bibinfo{year}{2017}).
\newblock \bibinfo{note}{ArXiv: 1703.00110}.

\bibitem{barkeshli_fractionalized_2016}
\bibinfo{author}{Barkeshli, M.}, \bibinfo{author}{Nayak, C.},
  \bibinfo{author}{Papic, Z.}, \bibinfo{author}{Young, A.} \&
  \bibinfo{author}{Zaletel, M.}
\newblock \bibinfo{title}{Fractionalized exciton {Fermi} surfaces and
  condensates in two-component quantized {Hall} states}.
\newblock \emph{\bibinfo{journal}{arXiv:1611.01171}}  (\bibinfo{year}{2016}).
\newblock \bibinfo{note}{ArXiv: 1611.01171}.

\bibitem{wang_one-dimensional_2013}
\bibinfo{author}{Wang, L.} \emph{et~al.}
\newblock \bibinfo{title}{One-{Dimensional} {Electrical} {Contact} to a
  {Two}-{Dimensional} {Material}}.
\newblock \emph{\bibinfo{journal}{Science}} \textbf{\bibinfo{volume}{342}},
  \bibinfo{pages}{614--617} (\bibinfo{year}{2013}).

\bibitem{dean_hofstadters_2013}
\bibinfo{author}{Dean, C.~R.} \emph{et~al.}
\newblock \bibinfo{title}{Hofstadter's butterfly and the fractal quantum {Hall}
  effect in moire superlattices}.
\newblock \emph{\bibinfo{journal}{Nature}} \textbf{\bibinfo{volume}{497}},
  \bibinfo{pages}{598--602} (\bibinfo{year}{2013}).

\bibitem{ponomarenko_cloning_2013}
\bibinfo{author}{Ponomarenko, L.~A.} \emph{et~al.}
\newblock \bibinfo{title}{Cloning of {Dirac} fermions in graphene
  superlattices}.
\newblock \emph{\bibinfo{journal}{Nature}} \textbf{\bibinfo{volume}{497}},
  \bibinfo{pages}{594--597} (\bibinfo{year}{2013}).

\bibitem{hunt_massive_2013}
\bibinfo{author}{Hunt, B.} \emph{et~al.}
\newblock \bibinfo{title}{Massive {Dirac} {Fermions} and {Hofstadter}
  {Butterfly} in a van der {Waals} {Heterostructure}}.
\newblock \emph{\bibinfo{journal}{Science}} \textbf{\bibinfo{volume}{340}},
  \bibinfo{pages}{1427--1430} (\bibinfo{year}{2013}).

\bibitem{ashoori_single-electron_1992}
\bibinfo{author}{Ashoori, R.~C.} \emph{et~al.}
\newblock \bibinfo{title}{Single-electron capacitance spectroscopy of discrete
  quantum levels}.
\newblock \emph{\bibinfo{journal}{Phys. Rev. Lett.}}
  \textbf{\bibinfo{volume}{68}}, \bibinfo{pages}{3088--3091}
  (\bibinfo{year}{1992}).

\bibitem{young_capacitance_2011}
\bibinfo{author}{Young, A.~F.} \& \bibinfo{author}{Levitov, L.~S.}
\newblock \bibinfo{title}{Capacitance of graphene bilayer as a probe of
  layer-specific properties}.
\newblock \emph{\bibinfo{journal}{Phys. Rev. B}} \textbf{\bibinfo{volume}{84}}
  (\bibinfo{year}{2011}).

\bibitem{jang_sharp_2016}
\bibinfo{author}{Jang, J.}, \bibinfo{author}{Hunt, B.~M.},
  \bibinfo{author}{Pfeiffer, L.~N.}, \bibinfo{author}{West, K.~W.} \&
  \bibinfo{author}{Ashoori, R.~C.}
\newblock \bibinfo{title}{Sharp tunnelling resonance from the vibrations of an
  electronic {Wigner} crystal}.
\newblock \emph{\bibinfo{journal}{Nature Physics}}
  \textbf{\bibinfo{volume}{advance online publication}} (\bibinfo{year}{2016}).

\bibitem{Jang_none_2017}
\bibinfo{author}{Jang, J.}
\newblock \bibinfo{title}{None}.
\newblock \emph{\bibinfo{journal}{Private Communication}}
  (\bibinfo{year}{2017}).

\bibitem{luryi_quantum_1988}
\bibinfo{author}{Luryi, S.}
\newblock \bibinfo{title}{Quantum capacitance devices}.
\newblock \emph{\bibinfo{journal}{Applied Physics Letters}}
  \textbf{\bibinfo{volume}{52}}, \bibinfo{pages}{501--503}
  (\bibinfo{year}{1988}).

\bibitem{goodall_capacitance_1985}
\bibinfo{author}{Goodall, R.~K.}, \bibinfo{author}{Higgins, R.~J.} \&
  \bibinfo{author}{Harrang, J.~P.}
\newblock \bibinfo{title}{Capacitance measurements of a quantized
  two-dimensional electron gas in the regime of the quantum {Hall} effect}.
\newblock \emph{\bibinfo{journal}{Phys. Rev. B}} \textbf{\bibinfo{volume}{31}},
  \bibinfo{pages}{6597--6608} (\bibinfo{year}{1985}).

\bibitem{geick_normal_1966}
\bibinfo{author}{Geick, R.}, \bibinfo{author}{Perry, C.~H.} \&
  \bibinfo{author}{Rupprecht, G.}
\newblock \bibinfo{title}{Normal {Modes} in {Hexagonal} {Boron} {Nitride}}.
\newblock \emph{\bibinfo{journal}{Physical Review}}
  \textbf{\bibinfo{volume}{146}}, \bibinfo{pages}{543--547}
  (\bibinfo{year}{1966}).

\bibitem{white_density_1992}
\bibinfo{author}{White, S.~R.}
\newblock \bibinfo{title}{Density matrix formulation for quantum
  renormalization groups}.
\newblock \emph{\bibinfo{journal}{Physical Review Letters}}
  \textbf{\bibinfo{volume}{69}}, \bibinfo{pages}{2863--2866}
  (\bibinfo{year}{1992}).

\bibitem{girvin_quantum_1987}
\bibinfo{editor}{Girvin, S.} \& \bibinfo{editor}{Prange, R.} (eds.)
  \emph{\bibinfo{title}{The {Quantum} {Hall} {Effect}}}
  (\bibinfo{publisher}{Springer-Verlag}, \bibinfo{year}{1987}).

\bibitem{jung_accurate_2014}
\bibinfo{author}{Jung, J.} \& \bibinfo{author}{MacDonald, A.~H.}
\newblock \bibinfo{title}{Accurate tight-binding models for the bands of
  bilayer graphene}.
\newblock \emph{\bibinfo{journal}{Phys. Rev. B}} \textbf{\bibinfo{volume}{89}}
  (\bibinfo{year}{2014}).

\bibitem{li_Entanglement_2008}
\bibinfo{author}{Li, H.} \& \bibinfo{author}{Haldane, F. D.~M.}
\newblock \bibinfo{title}{Entanglement {Spectrum} as a {Generalization} of
  {Entanglement} {Entropy}: {Identification} of {Topological} {Order} in
  {Non}-{Abelian} {Fractional} {Quantum} {Hall} {Effect} {States}}.
\newblock \emph{\bibinfo{journal}{Physical Review Letters}}
  \textbf{\bibinfo{volume}{101}}, \bibinfo{pages}{010504}
  (\bibinfo{year}{2008}).

\bibitem{kitaev_topological_2006}
\bibinfo{author}{Kitaev, A.} \& \bibinfo{author}{Preskill, J.}
\newblock \bibinfo{title}{Topological {Entanglement} {Entropy}}.
\newblock \emph{\bibinfo{journal}{Physical Review Letters}}
  \textbf{\bibinfo{volume}{96}}, \bibinfo{pages}{110404}
  (\bibinfo{year}{2006}).

\bibitem{morf_transition_1998}
\bibinfo{author}{Morf, R.~H.}
\newblock \bibinfo{title}{Transition from {Quantum} {Hall} to {Compressible}
  {States} in the {Second} {Landau} {Level}: {New} {Light} on the ν=5/2
  {Enigma}}.
\newblock \emph{\bibinfo{journal}{Physical Review Letters}}
  \textbf{\bibinfo{volume}{80}}, \bibinfo{pages}{1505--1508}
  (\bibinfo{year}{1998}).

\bibitem{rezayi_incompressible_2000}
\bibinfo{author}{Rezayi, E.~H.} \& \bibinfo{author}{Haldane, F. D.~M.}
\newblock \bibinfo{title}{Incompressible {Paired} {Hall} {State}, {Stripe}
  {Order}, and the {Composite} {Fermion} {Liquid} {Phase} in {Half}-{Filled}
  {Landau} {Levels}}.
\newblock \emph{\bibinfo{journal}{Phys. Rev. Lett.}}
  \textbf{\bibinfo{volume}{84}} (\bibinfo{year}{2000}).

\bibitem{geraedts_half-filled_2016}
\bibinfo{author}{Geraedts, S.~D.} \emph{et~al.}
\newblock \bibinfo{title}{The half-filled {Landau} level: {The} case for
  {Dirac} composite fermions}.
\newblock \emph{\bibinfo{journal}{Science}} \textbf{\bibinfo{volume}{352}},
  \bibinfo{pages}{197--201} (\bibinfo{year}{2016}).

\end{thebibliography}
\end{document}